%% file: main.tex
\documentclass[10pt, conference, compsocconf]{IEEEtran}

\usepackage{cite}
\usepackage{amsmath,amssymb,amsfonts}
\usepackage{algorithmic}

\usepackage{textcomp}
\usepackage{xcolor}
\usepackage{fancyhdr}
\usepackage[normalem]{ulem}
\usepackage[hyphens]{url}

\usepackage{graphicx}

\input{packages.tex}

\input{macros.tex}

\title{\name: \underline{N}oise-\underline{A}daptive \underline{S}earch for Robust Quantum Circuits}

\author{\IEEEauthorblockN{Hanrui Wang$^{1}$, Yongshan Ding$^{2}$, Jiaqi Gu$^{3}$, Zirui Li$^{5}$, Yujun Lin$^{1}$, David Z. Pan$^{3}$, Frederic T. Chong$^{4}$, Song Han$^{1}$}
\IEEEauthorblockA{$^{1}$Massachusetts Institute of Technology $^{2}$Yale University
$^{3}$University of Texas at Austin 
$^{4}$University of Chicago
$^{5}$SJTU\\
\normalfont{\texttt{\url{https://qmlsys.mit.edu}}}
}}

\begin{document}
\maketitle
\pagestyle{plain}
\thispagestyle{fancy}

\input{texts/0_abstract}

\input{texts/1_introduction}

\input{texts/2_background}

\input{texts/4_method}

\input{texts/5_implementation}

\input{texts/6_evaluation}

\input{texts/3_related}

\input{texts/7_conclusion}

\input{texts/acknowledgment}

\input{texts/artifact}

\bibliographystyle{IEEEtran}
\bibliography{main.bib}

\end{document}

%% file: packages.tex
\usepackage{color,xcolor}
\usepackage{xspace}
\usepackage{multirow}
\usepackage{booktabs}
\usepackage{lipsum}
\usepackage{soulutf8}

\usepackage{colortbl}

\let\labelindent\relax

\usepackage{enumitem}
\usepackage{braket}

\usepackage{amsmath}
\usepackage{amsfonts}
\usepackage{url}

\usepackage{afterpage}
\usepackage{ulem}
\usepackage[ruled,vlined]{algorithm2e}
\usepackage{setspace}

\usepackage{pifont}

\usepackage{titlesec}

\usepackage{hyperref}
\hypersetup{
    colorlinks=true,
    linkcolor=magenta,
    filecolor=magenta,      
    urlcolor=blue,
}

\usepackage{xcolor}
\usepackage{soul}



%% file: macros.tex
\definecolor{citecolor}{RGB}{34,139,34}
\definecolor{mydarkblue}{rgb}{0,0.08,1}
\definecolor{mydarkgreen}{rgb}{0.02,0.6,0.02}
\definecolor{mydarkred}{rgb}{0.8,0.02,0.02}
\definecolor{mydarkorange}{rgb}{0.40,0.2,0.02}
\definecolor{mypurple}{RGB}{111,0,255}
\definecolor{myred}{rgb}{1.0,0.0,0.0}
\definecolor{mygold}{rgb}{0.75,0.6,0.12}
\definecolor{myblue}{rgb}{0,0.2,0.8}
\definecolor{mydarkgray}{rgb}{0.,0.2,0.2}

\definecolor{lightred}{RGB}{255,235,235}
\definecolor{lightgreen}{RGB}{235,255,235}
\definecolor{lightblue}{RGB}{235,235,255}
\definecolor{lightcyan}{RGB}{235,255,255}
\definecolor{lightmagenta}{RGB}{255,235,255}
\definecolor{lightyellow}{RGB}{255,255,235}

\definecolor{qxkcolor}{RGB}{215,235,255}
\definecolor{softmaxcolor}{RGB}{230,235,255}
\definecolor{probxvcolor}{RGB}{255,255,235}

\definecolor{topkcolor}{RGB}{255,235,235}
\definecolor{zecolor}{RGB}{255,255,235}
\definecolor{dynacolor}{RGB}{235,255,255}

\definecolor{reviewcolor}{RGB}{0,0,200}

\renewcommand\footnotemark{}

\newcommand{\eg}{e.g., }
\newcommand{\etc}{{etc.}\xspace}

\newcommand{\name}{QuantumNAS\xspace}
\newcommand{\supercircuit}{SuperCircuit\xspace}
\newcommand{\subcircuit}{SubCircuit\xspace}

\newcommand{\nisq}{NISQ\xspace}
\newcommand{\supercircuits}{SuperCircuits\xspace}
\newcommand{\subcircuits}{SubCircuits\xspace}
\newcommand{\quantumengine}{\texttt{TorchQuantum}\xspace}
\newcommand{\water}{H$_2$O\xspace}
\newcommand{\lih}{LiH\xspace}
\newcommand{\chfour}{CH$_4$\xspace}
\newcommand{\behtwo}{BeH$_2$\xspace}
\newcommand{\htwo}{H$_2$\xspace}
\newcommand{\vs}{vs.\xspace}
\newcommand{\cnot}{\texttt{CNOT}}
\newcommand{\swap}{\texttt{SWAP}}
\newcommand{\rx}{\texttt{RX}}
\newcommand{\ry}{\texttt{RY}}
\newcommand{\rz}{\texttt{RZ}}
\newcommand{\rxyz}{\texttt{RXYZ}}
\newcommand{\uthree}{\texttt{U3}}
\newcommand{\cuthree}{\texttt{CU3}}
\newcommand{\uone}{\texttt{U1}}
\newcommand{\cuone}{\texttt{CU1}}
\newcommand{\hgate}{\texttt{H}}
\newcommand{\tgate}{\texttt{T}}
\newcommand{\sgate}{\texttt{S}}
\newcommand{\xx}{\texttt{XX}}
\newcommand{\zx}{\texttt{ZX}}
\newcommand{\zz}{\texttt{ZZ}}
\newcommand{\cz}{\texttt{CZ}}
\newcommand{\sqrtswap}{$\sqrt{\texttt{SWAP}}$}
\newcommand{\sqrth}{$\sqrt{\texttt{H}}$}

\newcounter{rlabelno}

%% file: texts/0_abstract.tex
\begin{abstract}
\label{sec:abstract}
Quantum noise is the key challenge in Noisy Intermediate-Scale Quantum (\nisq) computers. 
Previous work for mitigating noise has primarily focused on gate-level or pulse-level noise-adaptive compilation. 
However, limited research has explored a \textit{higher level of optimization} by making the quantum circuits themselves resilient to noise.

In this paper, we propose \name, a comprehensive framework for noise-adaptive co-search of the variational circuit and qubit mapping.
Variational quantum circuits are a promising approach for constructing quantum neural networks for machine learning and variational ansatzes for quantum simulation.
However, finding the best variational circuit and its optimal parameters is challenging due to the large design space and parameter training cost. We propose to \textit{decouple} the circuit search from parameter training by introducing a novel \textit{\supercircuit}. The \supercircuit is constructed with multiple layers of pre-defined parameterized gates (e.g., \texttt{U3} and \texttt{CU3}) and trained by iteratively sampling and updating the parameter subsets (\subcircuits) of it. It provides an accurate estimation of \subcircuits performance trained from scratch. Then we perform an evolutionary co-search of \subcircuit and its qubit mapping. The \subcircuit performance is estimated with parameters inherited from \supercircuit and simulated with real device noise models. Finally, we perform iterative gate pruning and finetuning to remove redundant gates in a fine-grained manner. 

Extensively evaluated with 12 quantum machine learning (QML) and variational quantum eigensolver (VQE) benchmarks on 14 quantum computers, \name significantly outperforms noise-unaware search, human, random, and existing noise-adaptive qubit mapping baselines. For QML tasks, \name is the first to demonstrate over 95\% 2-class, 85\% 4-class, and 32\% 10-class classification accuracy on real quantum computers. It also achieves the lowest eigenvalue for VQE tasks on \htwo, \water, \lih, \chfour, BeH$_{2}$ compared with UCCSD baselines. 
We also open-source the \href{https://github.com/mit-han-lab/torchquantum}{\quantumengine} library for fast training of parameterized quantum circuits to facilitate future research.
\end{abstract}

\begin{IEEEkeywords}
Quantum Computing; Quantum Noise; Variational Quantum Algorithm; Quantum Machine Learning; Neural Networks; Qubit Mapping; VQE; QNN
\end{IEEEkeywords}

%% file: texts/1_introduction.tex
\section{Introduction}
\label{sec:introduction}

Quantum Computing (QC) is a new computational paradigm that aims to address classically intractable problems with considerably higher efficiency and speed. 
It has been shown to have exponential or polynomial advantage in various domains such as cryptography~\cite{shor1999polynomial}, database search~\cite{grover1996fast}, chemistry~\cite{kandala2017hardware, peruzzo2014variational, cao2019quantum} and machine learning~\cite{biamonte2017quantum, harrow2009quantum, farhi2014quantum, lloyd2013quantum, rebentrost2014quantum}, \etc 
In the recent two decades, QC hardware has witnessed rapid progress by virtue of breakthroughs in physical implementation technologies. 

Despite the exciting advancements, we are still expected to reside in the Noisy Intermediate Scale Quantum (\nisq)~\cite{preskill2018quantum} stage for multiple years before entering the Fault-Tolerant era~\cite{gottesman2010introduction, bennett1996mixed}. 
In the NISQ era, quantum computers typically contain tens to hundreds of qubits, which are insufficient for quantum error correction. 
The qubits and quantum gates also suffer from high error rates of $10^{-3}$ to $10^{-2}$. 
Therefore, reducing quantum error is of pressing demand to close the gap between the requirements from the quantum algorithm side and available QC capacity from the hardware side.

\input{figtex/fig_teaser}

Many quantum system works have been proposed in recent years~\cite{zhang2021exploiting, 10.1145/3466752.3480072, li2021software, laodesigning, liu2021systematic, fatima2021faster, duckering2021orchestrated, 10.1145/3445814.3446750, tang2021cutqc, gokhale2020optimized, duckering2020virtualized, murali2020architecting, cheng2020accqoc, ding2020square, 10.1145/3373376.3378488, li2020towards}. Some of them focus on noise-adaptive quantum program compilation to mitigate the noise impact~\cite{das2021jigsaw, das2021adapt, 9499945, wu2021tilt, 9407180, 10.1145/3445814.3446706, 10.1145/3445814.3446743, 9251960, holmes2020nisq, murali2020software}.
Noise-adaptive qubit mapping~\cite{li2019tackling, murali2019noise, tannu2019not, tan2020optimal, tan2021optimal} aims to find the best mapping from logical qubits to physical qubits, which minimizes the gate error and SWAP insertion overhead. 
Noise-adaptive instruction scheduling and crosstalk mitigation techniques~\cite{ding2020systematic, murali2020software} aim to reduce the undesired inter-qubit interference and the circuit depth. 
However, those techniques only explore a small design space by optimizing the compilation process with a \emph{fixed} input quantum circuit. 
Limited research efforts have been made to explore how to improve the noise resilience of QC via a co-design strategy for searching, training, and compiling quantum circuits.

This work fills this blank by proposing \name, a \underline{n}oise-\underline{a}daptive quantum circuit and qubit mapping co-\underline{s}earch
framework to find the most robust quantum circuit and corresponding qubit mapping tailored for a given task on the target quantum device as in Figure~\ref{fig:teaser}. 
We study variational quantum circuits (trainable circuits with parameterized quantum gates) since they provide unique opportunities to alter circuit structures while performing the same functionality.

\input{figtex/fig_sim_vs_real}

First, we are strongly motivated by the significant impacts of quantum noise on performance. 
In Figure~\ref{fig:sim_vs_real}, we show the accuracy of MNIST 4-class image classification simulated by the noise-free simulator and measured on the real IBMQ-Yorktown quantum computer. 
\emph{Key observations}: (1) More parameters increase the model capacity, thus increasing noise-free simulation accuracy. 
Nevertheless, more parameters mean more gates, which introduces more noise, and the accumulated noise quickly offsets the capacity benefit. 
As a result, the measured accuracy peaks at 45 parameters. 
(2) To make things worse, quantum noise exacerbates the performance variance. The measured accuracy variance under the same \#parameters is much higher than that of noise-free, \eg [25\%, 59\%] \vs [67\%, 77\%] under 45 parameters. 
The observations both call for the noise-adaptive search for the most robust circuit.

One major challenge for this noise-adaptive search is the algorithmic scalability issue.
It is almost intractable to solve the two-level optimization problem (for quantum circuit and qubit mapping) via iterative circuit sampling, parameter training, and evaluation in the large design space.
To address this, we propose to \emph{decouple the training and search} by introducing a novel \emph{\supercircuit}-based search approach (Figure~\ref{fig:teaser}). 
We first construct a SuperCircuit by stacking a sufficient number of layers of pre-defined parameterized gates to cover a large design space. Then, we train the \supercircuit by sampling and updating the parameter subsets (\subcircuits) from the \supercircuit. 
The performance of a \subcircuit with inherited parameters from the \supercircuit can provide a reliable relative performance estimation for the individual \subcircuit trained from scratch. 
In this way, we only pay the training cost \emph{once} but can evaluate \emph{all} the \subcircuits fast and efficiently. Hence, the search cost is significantly reduced. 

\input{figtex/fig_curve}
Furthermore, we perform an evolutionary co-search with noise information in the loop to find the most robust quantum circuit and qubit mapping jointly. 
In each iteration, the evolution engine samples a population of \subcircuit and qubit mapping pairs. 
Then the performance of each sampled \subcircuit can be evaluated by an estimator on two types of backends: a \emph{noise-aware simulator} or a \emph{real quantum hardware}. 
The estimator takes the inherited parameters from the \supercircuit and assigns them to the \subcircuit. With a noise-aware simulator backend, the performance is evaluated with direct noise classical simulation with a realistic device noise model. Alternatively, we can replace the simulator with real quantum hardware. The requirement for such evaluation is \emph{no harder than} any common variational quantum algorithms. 
After multiple evolutionary search iterations,
we can obtain a pair of robust circuit and qubit mapping and then train the parameters from scratch. SuperCircuit-based search is inspired by the supernet method in classical ML model training~\cite{pham2018efficient, guo2020single, cai2019once, wang2020hat, wang2020efficient, tang2020searching}. However, we have five major differences: (1) The SuperCircuit is more general than the ML model and can be applied to various parameterized quantum algorithms such as VQE; (2) We co-search circuit with its qubit mapping; (3) Our search is aware of quantum noise to improve robustness; (4) We propose novel \emph{front sampling} and \emph{restricted sampling} specialized for quantum circuits. (5) We experimentally demonstrate the feasibility of training circuits on real QC with on-device gradient computation.

Finally, on top of the searched circuit and qubit mapping, we further propose a fine-grained \emph{pruning} technique to remove redundant parameters and gates and \emph{finetuning} to recover the performance. 
We end up with a slimmed circuit with similar noise-free performance but fewer noise sources, which in return improves the final measured performance. 

Overall, \name can mitigate the impact of quantum noise and delays the accuracy peak as shown in Figure~\ref{fig:curve}.
The contributions of \name are five-fold: \ding{202} \textbf{Noise-Adaptive Quantum Circuit \& Qubit Mapping Co-Search} to enable noise-resilient QC. \ding{203} \textbf{\supercircuit-based Efficient Search Flow}: we propose a scalable quantum circuit search method based on \supercircuit. Front sampling and restricted sampling are proposed for efficient exploration and stable optimization in the huge design space. \ding{204} \textbf{Iterative Quantum Pruning} is introduced to remove redundant quantum gates in a fine-grained manner. \ding{205} \textbf{Extensive Real QC Evaluations}: we extensively evaluate \name with 12 benchmarks in QML and VQE on 14 quantum computers, observing significant improvements over baselines. \ding{206} \textbf{Open-Source QC Library}: To facilitate future research in QML and variational quantum simulation, we release \href{https://github.com/mit-han-lab/torchquantum}{\quantumengine}, a PyTorch-based GPU-accelerated library to enable fast training of parameterized quantum circuits (over 200$\times$ faster than the PennyLane~\cite{bergholm2018pennylane}). It also supports push-the-button deployment of trained circuits on real quantum devices.

%% file: figtex/fig_teaser.tex
\begin{figure}[t]
    \centering
    \includegraphics[width=\columnwidth]{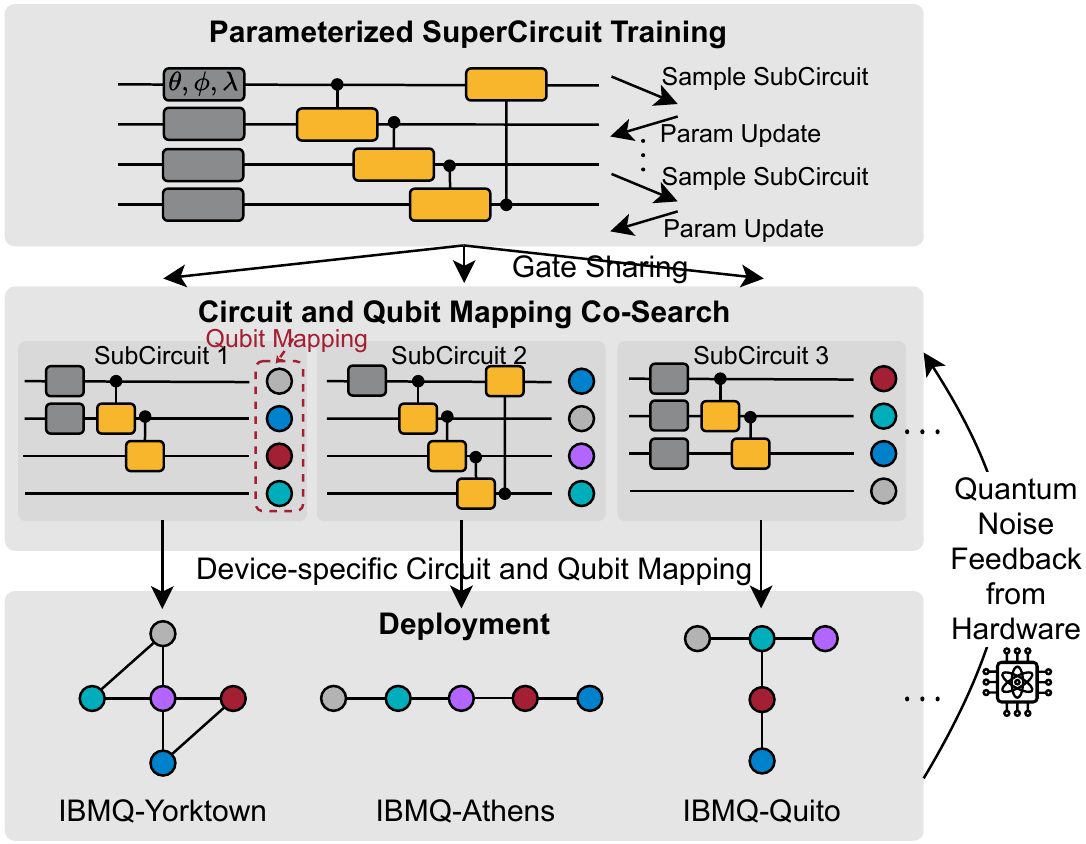}
    \vspace{-20pt}
    \caption{Noise-adaptive circuit and qubit mapping co-search. 
    A gate-sharing SuperCircuit that contains numerous parameter subsets (SubCircuit) is firstly trained.
    Then we perform an evolutionary search with the quantum noise feedback to find the most robust circuit and qubit mapping. }
    \label{fig:teaser}
    \vspace{-10pt}
\end{figure}

%% file: figtex/fig_sim_vs_real.tex
\begin{figure}[t]
    \centering
    \includegraphics[width=\columnwidth]{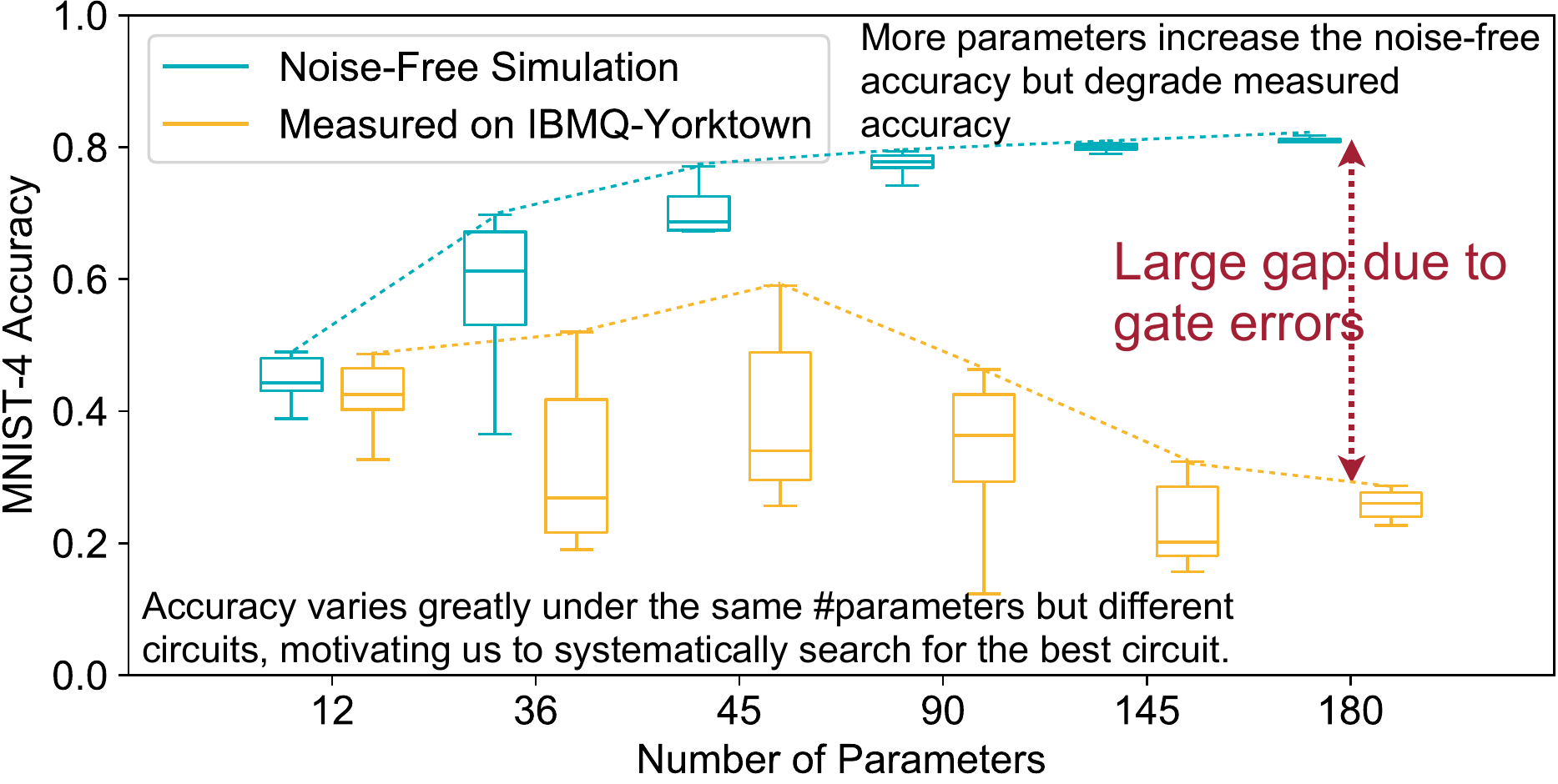}
    \caption{MNIST-4 
    on noise-free simulator / real QC. 
    More parameters increase the noise-free accuracy but degrade measured accuracy due to larger gate errors. Accuracy varies greatly under the same \#parameters but different circuits, motivating us to search for the best circuit systematically.
    }
    \label{fig:sim_vs_real}
    \vspace{-10pt}
\end{figure}

%% file: figtex/fig_curve.tex
\begin{figure}[t]
    \centering
    \includegraphics[width=\columnwidth]{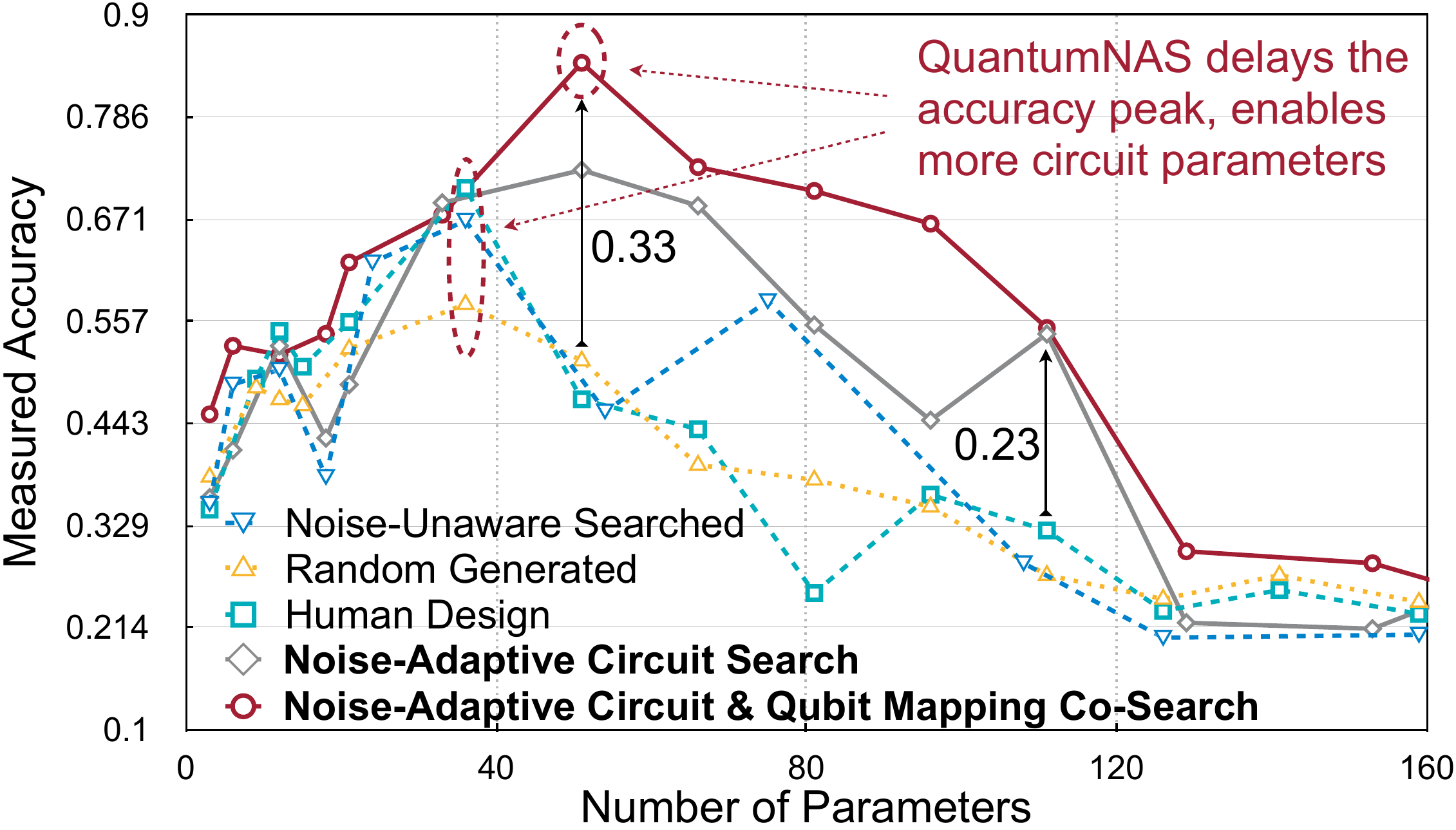}
     \vspace{-20pt}
    \caption{Accuracy \vs \#parameters of multiple methods. 
    The accuracy of conventional designs quickly saturates then drops. 
    \name mitigates the quantum noise and delays the peak of the curve, allowing larger model capacity and higher accuracy (up to 33\% higher).
    }
    \label{fig:curve}
     \vspace{-13pt}
\end{figure}

%% file: texts/2_background.tex
\section{Background and Motivation}
\label{sec:background}

\subsection{Quantum Basics}
\noindent\textbf{Qubits.}
The power of quantum computation stems from its fundamentally unique way of storing and manipulating information \cite{nielsen2002quantum,ding2020quantum}. 
Unlike a conventional bit, a quantum bit (\emph{qubit}) can be in a linear combination of the two basis states 0 and 1: $\ket{\psi} = \alpha\ket{0} + \beta\ket{1},$ for $\alpha,\beta\in\mathbb{C},$ satisfying $|\alpha|^2+|\beta|^2=1$. The ability to create a ``superposition'' of basis states allows us to use an $n$-qubit system to represent a linear combination of $2^n$ basis states. 
In contrast, a classical $n$-bit register can only store one of the $2^n$ states.

\noindent\textbf{Quantum Circuits.} 
To perform computation on a quantum system, we manipulate the qubits' state by applying a \emph{quantum circuit}. 
A quantum circuit consists of a sequence of operations called \emph{quantum gates}, which take one quantum state to another through unitary transformations, i.e., $\ket{\psi} \rightarrow U\ket{\psi}$, where $U$ is a unitary matrix. Results of a quantum circuit are obtained by qubit readout operations called \emph{measurements}, which collapse a qubit state $\ket{\psi}$ to either $\ket{0}$ or $\ket{1}$ probabilistically according to the amplitudes $\alpha$ and $\beta$.
Finding the best quantum circuits for a computational task is non-trivial \cite{ding2020quantum} -- a realistic, robust quantum circuit must: (1) faithfully express the desired transformation; (2) complete in an efficient number of steps; (3) be able to be implemented on hardware with reasonable fidelity.

\noindent\textbf{Operational Noises.}
In real QC, 
errors occur due to imperfect control signals, unwanted interactions between qubits, or interference from the environment \cite{krantz2019quantum, bruzewicz2019trapped, magesan2012characterizing}. Thus, 
qubits undergo \emph{decoherence error}  over time, and quantum gates introduce \emph{operation errors} (e.g., coherent/stochastic errors) into the system. 
These systems need to be characterized\cite{magesan2012characterizing} and calibrated\cite{ibm_2021} frequently to mitigate noise impacts.
So noise-adaptive techniques in QC algorithms, circuits, and devices are critical for operating quantum computers.

\subsection{Variational Quantum Circuits}
A variational circuit is a trainable quantum circuit where its quantum gates are parameterized (e.g., by angles in quantum rotation gates). The parameterized quantum circuit $\Phi(x,\theta)$ is used to prepare a variational quantum state: $\ket{\psi(x, \theta)} = \Phi(x, \theta)\ket{0\dotsc 0}$, where $x$ is the input data related to the computation and $\theta$ is a set of free variables for adaptive optimizations. Variational methods have shown huge potentials in applications such as quantum ML\cite{wittek2014quantum, biamonte2017quantum,schuld2018supervised,benedetti2019parameterized}, numerical analysis\cite{lloyd2014quantum, lloyd2016quantum}, quantum simulation \cite{peruzzo2014variational, kandala2017hardware, kokail2019self, mcclean2016theory, o2016scalable}, and optimizations\cite{moll2018quantum,farhi2014quantum}.

Typically, the training of variational circuits is performed by first selecting a hand-designed circuit for a computational task and, secondly, finding an optimal set of parameters for the circuit via a hybrid quantum-classical optimization procedure. The optimization is usually an iterative process to search for the best candidates for the parameters in $\Phi(x,\theta)$. 
Whether a variational quantum algorithm is successful depends on how well the circuit can be trained. For example, ``barren plateau''\cite{mcclean2018barren} is a phenomenon when the cost function landscape is flat, making a variational circuit untrainable with gradient-based optimizations.

\input{figtex/fig_qml_vqe}
\input{figtex/fig_overview}

Quantum Neural Network (QNN) is a promising application of variational quantum circuits~\cite{abbas2021power}. Figure~\ref{fig:qml_vqe} shows the example circuits we used for QML (QNN) and VQE. For QML tasks such as image classification, we first encode the pixels using rotation gates and then use parameterized trainable quantum gates to process the information. We measure the qubits on Z-basis to obtain classical values, then compute Softmax of those values to get the probability for each class. For VQE, the parameterized circuit is used for state preparation, and the measurement part is constructed according to the molecule. We prepare for the state multiple times for measurements on different qubits and bases, multiply expectation values of qubits, and perform weighted sum. The final result is the expectation value for the ground state energy of the molecule. The parameters can be trained with backpropagation, in which we compute the derivative of each parameter ($\theta_{i}$) on loss function ($L$) and update the parameters with a learning rate $\alpha$, $\hat{\theta_{i}} = \theta_{i} - \alpha \frac{\partial L}{\partial \theta_{i}}$.

%% file: figtex/fig_qml_vqe.tex
\begin{figure}[t]
    \centering
    \includegraphics[width=\columnwidth]{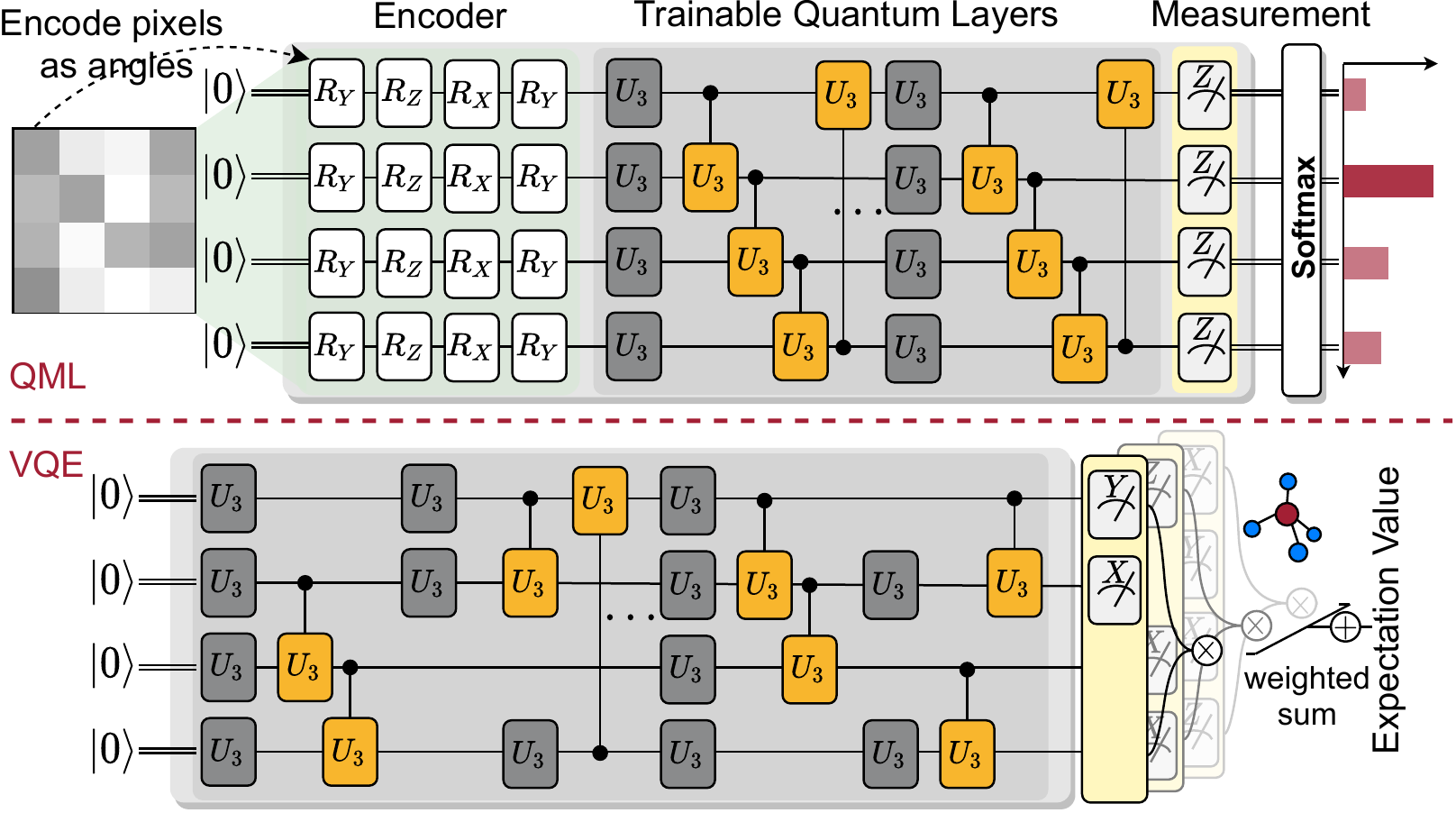}
    \vspace{-20pt}
    \caption{Example circuits for QML and VQE tasks.
    }
    \label{fig:qml_vqe}
    \vspace{-15pt}
\end{figure}

%% file: figtex/fig_overview.tex
\begin{figure*}[t]
    \centering
    \includegraphics[width=\textwidth]{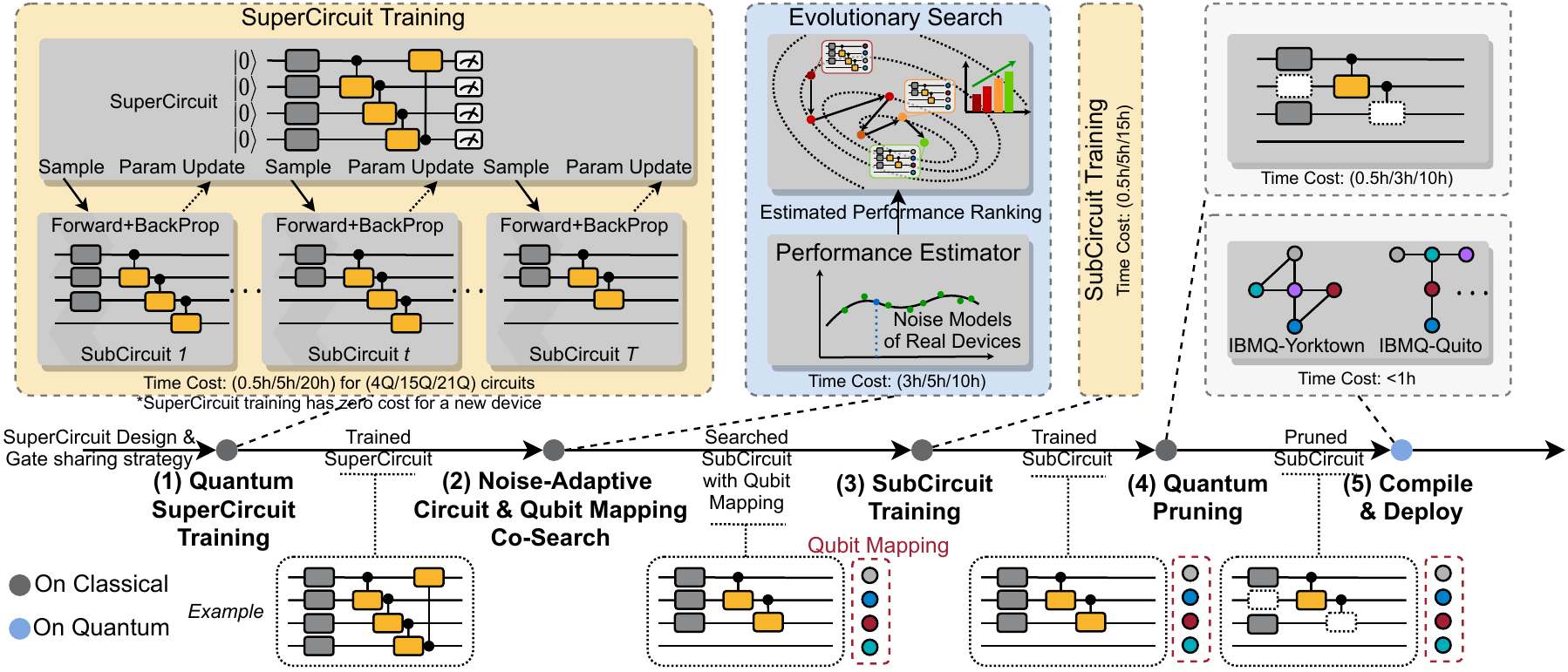}
    \vspace{-20pt}
    \caption{\name Overview. (1) A \supercircuit is trained by iteratively sampling and updating parameter subsets (\subcircuits). 
    The parameters from \supercircuit and the simulator with practical noise models can provide an accurate final performance ranking estimation of \subcircuits. 
    (2) Evolutionary co-search for circuit and qubit mapping pair of best estimated performance (lowest validation loss/eigenvalue for QML/VQE). (3) Train the searched SubCircuit. (4) Iterative pruning and finetuning to remove redundant gates. (5) Compile and deploy on real devices.
    }
    \label{fig:overview}
    \vspace{-10pt}
\end{figure*}

%% file: texts/4_method.tex
\section{Noise-Adaptive \name}
\label{sec:method}

\subsection{Overview}
Figure~\ref{fig:overview} shows \name overview, time cost, and a simple example. Firstly, a \supercircuit is trained as a fast estimation of \subcircuits performance ranking. We show several sampled example SubCircuits in the diagram. \emph{Front sampling} and \emph{restricted sampling} are proposed to promote the reliability of estimations. Then a noise-adaptive evolutionary co-search is performed to find the best circuit and qubit mapping pair. A performance estimator is employed to provide fast and accurate feedback to the evolution engine. 
Redundant gates with small parameter magnitude are further pruned from the searched circuit. The pruned circuit is finally compiled and deployed on real quantum devices.

\subsection{\supercircuit Construction and Training}
It is critical to encompass a large design space to include the most robust circuit.
However, training all candidate circuits, evaluating their final performance, and selecting the best one is too costly. 
We thus propose \supercircuit to evaluate each circuit in the design space (\subcircuit) \emph{without} fully training it. Since we only need to find the best circuit, \emph{relative performance} is sufficient and can be estimated by the \supercircuit.

With pre-specified basis gates and design space, the SuperCircuit is defined as the circuit with the largest number of gates in the space, whose parameters are trained by iteratively sampling and updating a subset of gates/parameters (\subcircuit). SuperCircuit contains multiple blocks, each with several layers of parameterized gates.  A \subcircuit is a subset of the \supercircuit that can have a different number of blocks and gates inside blocks. Figure~\ref{fig:sampling} shows one block of \uone+\cuone~space containing one \uone~layer and one \cuone~layer. The \supercircuit contains all gates, while the \subcircuit only contains gates with solid lines. In one \supercircuit training step, we sample a \subcircuit and only compute gradients using the \subcircuit and update that subset of parameters of \supercircuit. Intuitively, training a \supercircuit is simultaneously training all \subcircuits in the design space. All gates in SuperCircuit are single/two-qubit gates, thus are local interactions in variational ansatz.

\input{figtex/fig_sampling}

\input{figtex/fig_gene}

\supercircuit aims to facilitate the low-cost evaluation of \subcircuits in the design space. Given one \subcircuit, it is sufficient to inherit the gate parameters from the \supercircuit and then perform evaluation \emph{without} training. That provides an accurate estimation of the relative performance of the \subcircuit.
Since the next stage is derivative-free
optimization such as evolutionary search, using relative performance between \subcircuits is sufficient to find the best one.
In addition, \supercircuit can be reused for new devices or when noise changes. Thus, we only need to pay the noise-free \supercircuit training cost for \emph{once} but can use it for \emph{all} devices.
The number of circuits run for na\"ive search is $ N_{device} \times N_{search} \times (N_{train} + N_{eval}) $; while that for \supercircuit search is $ 1 \times N_{train} + N_{device} \times N_{search} \times N_{eval} $.
The overall search cost is significantly reduced by around $N_{device} \times N_{search}$ times which is $10 \times 1600=16,000$ in our setting. $N_{device}$ means  \#quantum devices to execute the circuit. $N_{search}$ means \#evaluated circuits during search. $N_{train}$/$N_{eval}$ means \#circuit running iterations in training/evaluation.

\input{figtex/fig_evolution}

A critical challenge in sampling-based \supercircuit training is the large variance. 
Na\"ive random sampling often causes severe trainability issues due to intractable sampling variance from drastic \subcircuit change, leading to unreliable relative performance estimation.
To address this, we propose \emph{front sampling} and \emph{restricted sampling}.

\textbf{Front Sampling.} In front sampling, only the subsets with the \emph{several front blocks and front gates} can be sampled. For instance, if the subset contains three blocks, then blocks 0, 1, 2 will be sampled. Inside a block, if two gates are sampled in a layer, then the gates on qubits 0 and 1 will be sampled. Figure~\ref{fig:sampling} shows several valid cases of front sampling. In the leftmost example, only the front two U$_1$ gates and the front three CU$_1$ gates are sampled. So in that step, only those five parameters are updated. Front sampling helps improve \supercircuit trainability as \subcircuits share the parameters of front blocks and gates.

\textbf{Restricted Sampling} is another essential technique we propose to boost training stability. 
We prevent the sampled \subcircuits from changing dramatically between two steps by constraining the maximum number of different layers.
Therefore, the training process is stabilized as the sampling variance is under control. 
As in Figure~\ref{fig:gene}, the upper path is unrestricted sampling where the two \subcircuits differ by 5 layers. 
In the bottom path, restricted sampling limits the layer differences to 3, bringing better cross-step consistency.

\subsection{Noise-Adaptive Evolutionary Co-Search}
\supercircuit provides highly efficient relative performance estimations.
We adopt a derivative-free optimization to explore the joint space of circuit and qubit mapping.

\textbf{Evolutionary Search.} Genetic algorithm is employed in which the gene vector encodes circuit and qubit mapping. Each element in the circuit sub-gene represents the circuit width (\#gates) in the layer. One additional gene sets the circuit depth (\#blocks). Front sampling is also applied here. The qubit mapping sub-gene encodes the mapping between logical and physical qubits. 
We concatenate circuit and qubit mapping sub-genes as the pair's gene.

\input{figtex/fig_corr}
\input{figtex/fig_scatter}

The evolution engine keeps a population of pairs and searches for high-performance candidates. In one iteration, it first evaluates all pairs by querying a performance estimator and selects multiple pairs with the highest performance (the lowest loss/eigenvalue for QML/VQE) as the parent population. 
Then mutation and crossover are conducted to generate the new population as in Figure~\ref{fig:evolution}. Mutation randomly alters several genes with a pre-defined probability. Crossover first selects two parent samples from the parent population; and then generates a new sample, each gene of which is randomly selected from one of the two parent samples. If the qubit mapping sub-gene contains a repeated qubit, we will replace the repeated one with the first unused qubit.
The new population is the ensemble of parent population, mutations, and crossovers. Then we sort the new population and select the ones with the highest performance as parents and enter the next iteration. The population of the very first iteration is from random sampling. Population size across iterations remains the same. For QML, we use validation set loss as the indicator. The lower the validation loss, the higher the final accuracy.

\textbf{Performance Estimator.} Ideally, the performance of circuit-qubit mapping pairs is directly evaluated on real quantum devices, which, however, could be extremely slow due to limited resources and queuing. Therefore we apply an estimator to provide \emph{fast relative performance with noise}.
It takes the query pairs from the evolution engine as inputs. Then, it inherits the gate parameters of searched \subcircuit from \supercircuit and sets the searched qubit mapping as the initial mapping of the compiler. There are two ways of estimation. 
One way is to use a simulator with a noise model from real devices. Noise models are from calibrations such as randomized benchmarking performed by the IBMQ team. They contain coherence (depolarizing), decoherence (thermal relaxation), and SPAM (readout) errors. The models are updated around twice a day and can be directly accessed with Qiskit API; the second is to use a noise-free simulator and compute the overall success rate with the product of success rates of all gates. Then the augmented loss will be noise-free simulated loss divided by calculated success rate:
${ r_{overall} = \prod_{i} r_{gate_{i}}, l_{augmented} = \frac{l_{noise\_free}}{r_{overall}}}$, where $r$ is the success rate, and $l$ is the loss.
The first method is more accurate but slower, while the second is less accurate but faster. 
Therefore, in \name, small circuits ($\leq$10 Qubits) apply the first method; large circuits apply the second method.

\input{figtex/fig_pruning}

The estimator has two approximations. 
The first uses the performance of one \subcircuit with \emph{inherited} parameters to estimate the performance of the same \subcircuit with parameters \emph{trained from scratch.} 
The second uses the simulation results, either with noise model or success rate, to estimate the performance on real devices. 
Since we only care about relative performance, the two-level approximation still maintains enough reliability for the search engine. Figure~\ref{fig:corr} shows the effectiveness of the first approximation with five tasks in two design spaces. For each point, the x-axis value is the performance (loss) with inherited parameters from SuperCircuit; the y-axis value is that with parameters trained from scratch. 
The average Spearman's correlation score is 0.75, showing \emph{strong positive correlations} thus accurate relative performance. 
Figure~\ref{fig:scatter} further shows the final estimated loss and the real loss for MNIST-4 on IBMQ-Yorktown. The correlation between them is 0.76, a strong positive correlation. Thus, the estimated performance is reliable enough to search for the best circuit-mapping pair.

\subsection{Iterative Quantum Pruning}
We further propose to remove redundant quantum gates to reduce the noise, inspired by the classical NN pruning~\cite{han2015deep, wang2021spatten, zhang2020sparch, pmlr-v123-yan20a, He_2018_ECCV, Wang_2020_CVPR} and pruning for noise-robust analog neurocomputing~\cite{gu2020fftonn, gu2021squeezelight,gu2021mixedtrain,xia2017reram,yuan2021reram}. 
The motivations are three-fold.
First, the sub-optimality of the evolutionary search stage leaves room for further optimization of the searched circuit by reducing the number of gates.
Second, even with the same circuit, there exist multiple parameter sets to achieve similar noise-free performance. 
Some sets contain more parameters with a magnitude close to zero, which can be safely removed with iterative pruning and finetuning.
Third, some gates, such as \uthree, contain multiple parameters. Partially removing the parameters can also bring benefits. \#compiled gates of U3($\theta,\phi,\lambda$),  U3($\mathbf{0},\phi,\lambda$), U3($\theta, \phi, \mathbf{0}$),
U3($\theta, \mathbf{0}, \lambda$), U3($\theta, \mathbf{0}, \mathbf{0}$), U3($\mathbf{0}, \phi, \mathbf{0}$) and U3($\mathbf{0}, \mathbf{0}, \lambda$) are 5, 1, 4, 4, 4, 1, 1, respectively. 
Therefore, having one or two parameters as zeros in the \uthree~gates can reduce up to 80\% gates compiled to the basis gate set (\cnot, \texttt{SX}, \rz).

Therefore, we propose iterative pruning to remove the circuit parameters in a fine-grained manner, shown in Figure~\ref{fig:pruning}. 
Specifically, we first train the searched circuit from scratch to convergence. 
We rank all the normalized rotation angles $(\theta,\phi,\lambda)\in[-\pi,\pi)$ and remove part of angles that are closest to 0$^{\circ}$.
Then we finetune the rest parameters to recover the accuracy. 
We iteratively increase the pruning ratio and finetune the circuit parameters until achieving the desired ratio.
In practice, we adopt polynomial pruning ratio decay~\cite{zhu2017prune}: $r_{now} = r_{final} + (r_{initial} - r_{final})  \left(1 - \frac{s_{now} - s_{begin}}{s_{end} - s_{begin}}\right) ^ 3$ where $r$ is pruning ratio, and $s$ is training step. For final pruning ratio selection, we make sure that the noise-free simulation performance is not degraded compared with the un-pruned circuit. Thus, due to fewer gates and fewer noise sources after compilation, the accuracy of the circuit can be further increased by up to 9\%.

%% file: figtex/fig_sampling.tex
\begin{figure*}[t]
    \centering
    \includegraphics[width=\textwidth]{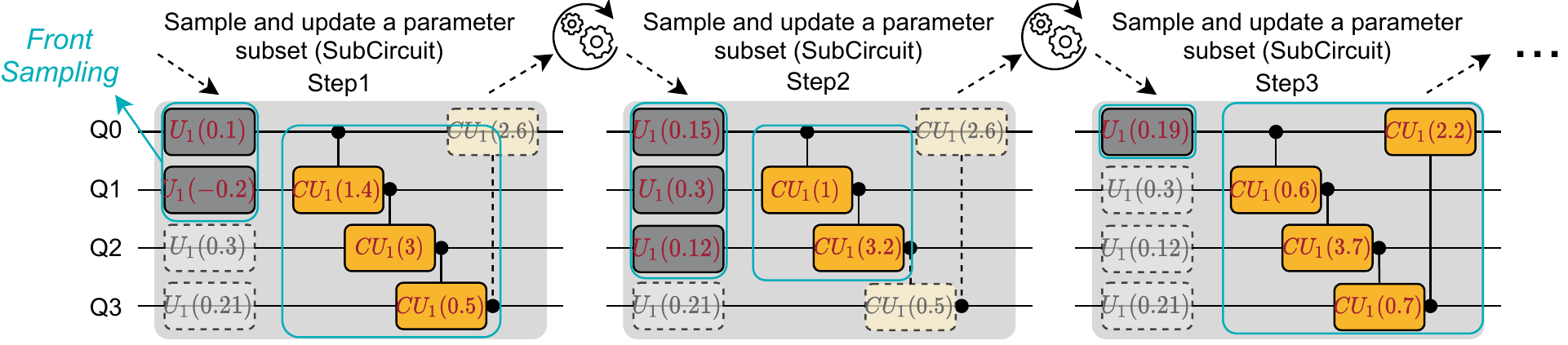}
    \vspace{-20pt}
    \caption{SuperCircuit training. At each step, a subset of SuperCircuit parameters (SubCircuit) is sampled and then updated.
    }
    \label{fig:sampling}
    \vspace{-15pt}
\end{figure*}

%% file: figtex/fig_gene.tex
\begin{figure}[t]
    \centering
    \includegraphics[width=\columnwidth]{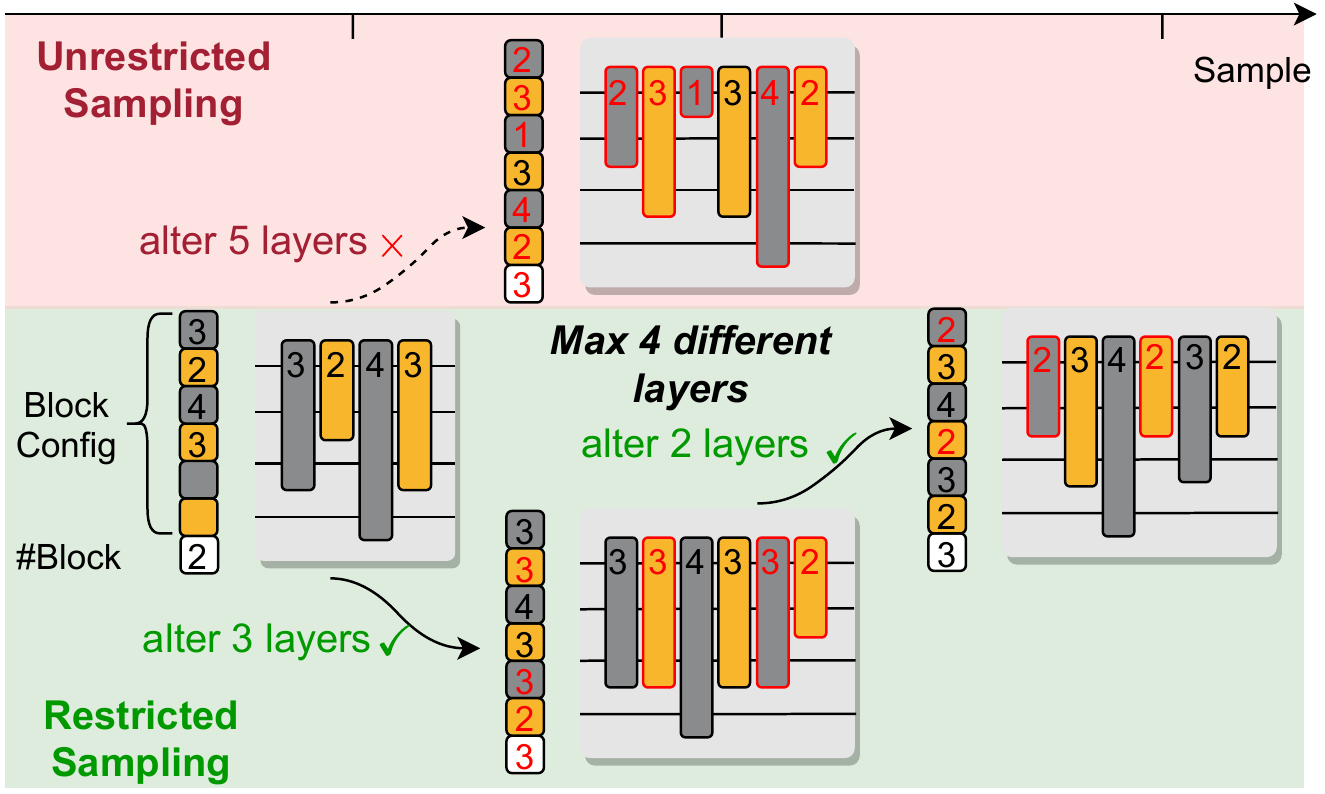}
    \vspace{-20pt}
    \caption{Restricted sampling constrains \#layers that are different between two steps. It improves \subcircuit consistency thus stabilizes the \supercircuit training process.}
    \label{fig:gene}
    \vspace{-10pt}
\end{figure}

%% file: figtex/fig_evolution.tex
\begin{figure}[t]
    \centering
    \includegraphics[width=\columnwidth]{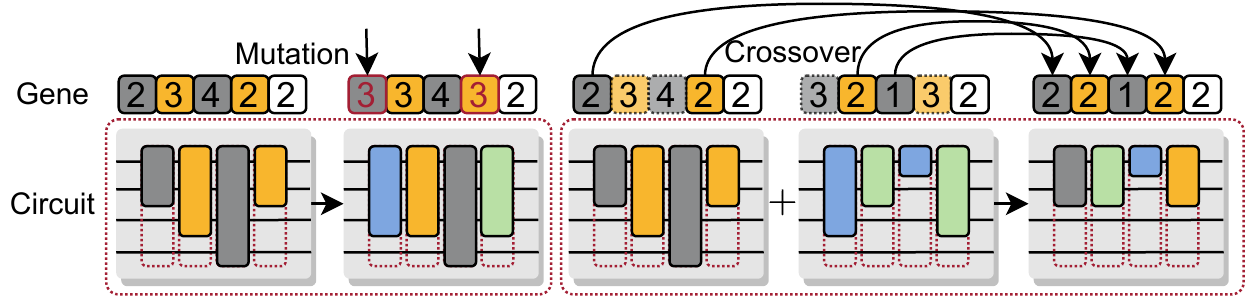}
    \vspace{-20pt}
    \caption{\subcircuit mutation and crossover in evolutionary search.}
    \vspace{-10pt}
    \label{fig:evolution}
\end{figure}

%% file: figtex/fig_corr.tex
\begin{figure}[t]
    \centering
    \includegraphics[width=\columnwidth]{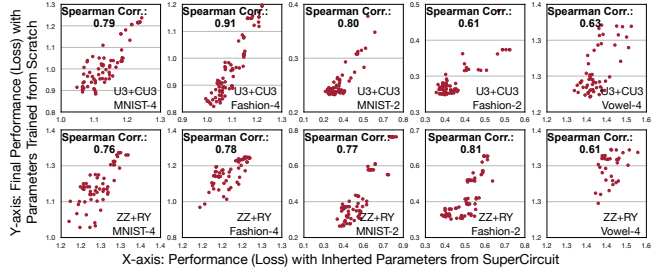}
    \vspace{-20pt}
    \caption{Strong correlation between performance with inherited parameters from SuperCircuit and parameters trained from scratch.}
    \label{fig:corr}
    \vspace{-10pt}
\end{figure}

%% file: figtex/fig_scatter.tex
\begin{figure}[t]
    \centering
    \includegraphics[width=\columnwidth]{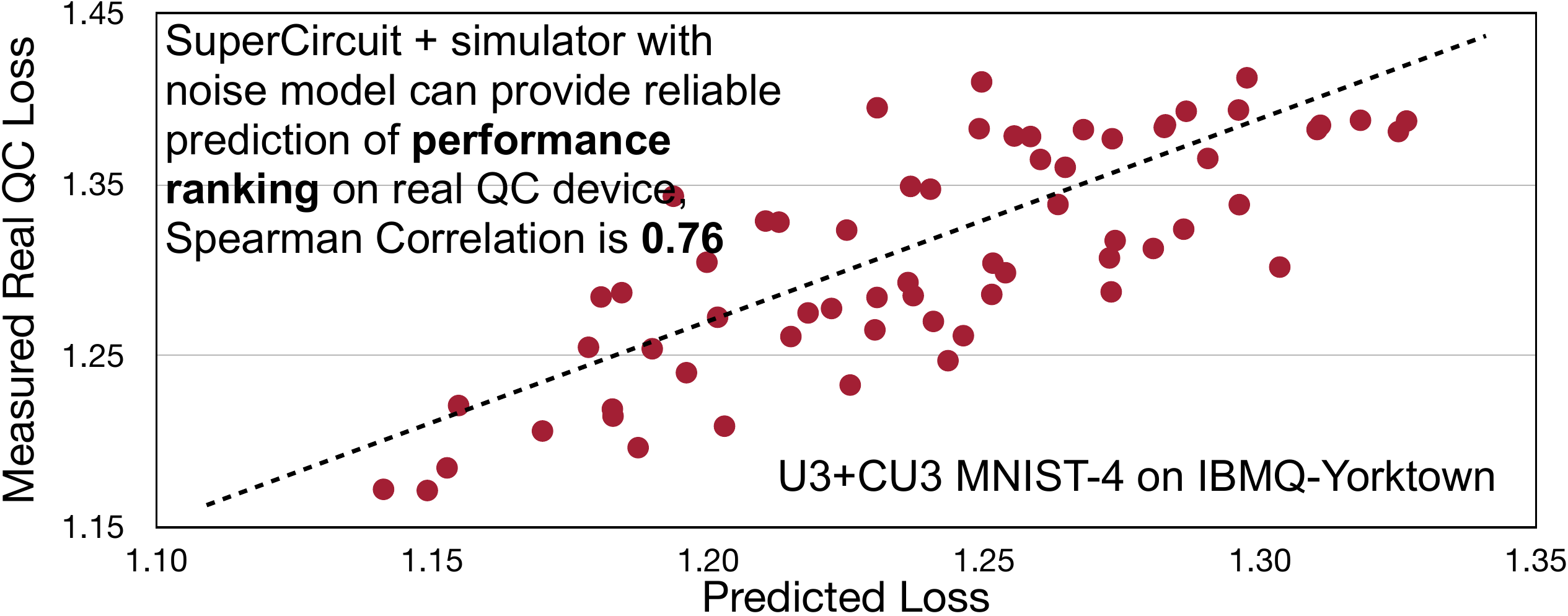}
    \vspace{-20pt}
    \caption{Reliability of the performance estimator. The estimated loss can accurately indicate the final loss.
    }
    \label{fig:scatter}
    \vspace{-15pt}
\end{figure}

%% file: figtex/fig_pruning.tex
\begin{figure}[t]
    \centering
    \includegraphics[width=\columnwidth]{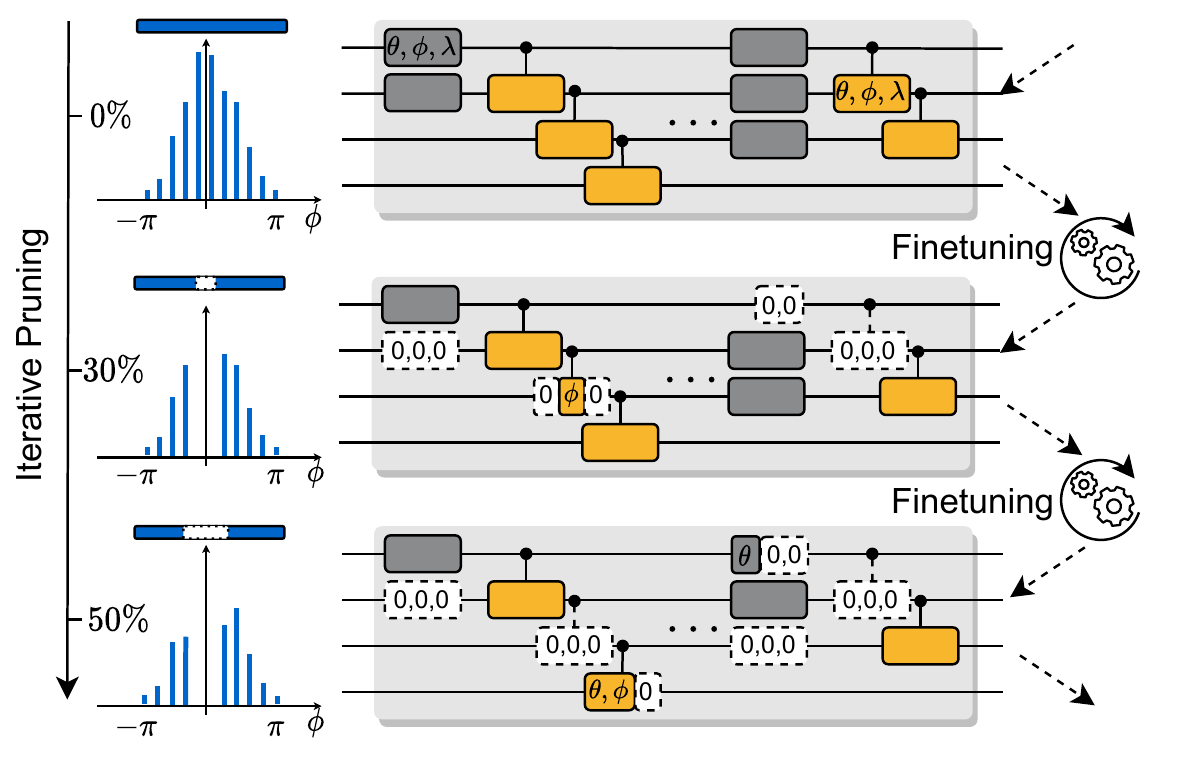}
    \vspace{-20pt}
    \caption{Iterative pruning and finetuning remove small-magnitude parameters. 
    thus reducing the number of compiled gates.
    }
    \label{fig:pruning}
    \vspace{-10pt}
    
\end{figure}

%% file: texts/5_implementation.tex
\input{figtex/fig_quantum_engine}
\subsection{\quantumengine Library}
\label{sec:implementation}

To accelerate parameterized quantum circuit training in this work, we build a PyTorch library named \quantumengine. Its APIs are implemented similarly to existing operations in PyTorch. So it makes quantum circuit construction as easy as a standard neural network model. 
It supports all common quantum gates. 
The state vector and unitary matrix of each gate are implemented with a native \texttt{torch.Tensor} data type. 
The simulations are achieved with complex-valued differentiable matrix multiplication operators such as \texttt{torch.bmm}. 

There exists several QNN training frameworks such as PennyLane~\cite{bergholm2018pennylane}. The major advantage is its flexible interfaces to various frameworks such as PyTorch, TensorFlow, JAX, Keras, and NumPy. However, the operations are not implemented using native ones in those frameworks, so off-the-shelf optimizations of those frameworks cannot be used. Moreover, Pennylane can only use \emph{parameter shift} to obtain gradients, which is inherently sequential, so no parallelization on batch and gate dimension can be achieved.
Compared with Pennylane, QuantumEngine has several unique advantages: (1) It supports both dynamic and static computational graphs. 
Dynamic mode simulates each gate individually, so the state vector after each gate can be obtained for easy debugging (statevector simulation).
Static mode optimizes tensor network simulation by fusing unitary of multiple gates before applying to the state vector, reducing the computation amount (tensor network simulation).
(2) It supports both parameter shift and back-propagation training. In back-propagation mode, training is highly parallelized. PennyLane only supports parameter shift and processes batch with inefficient 'For' loop. 
(3) All simulations can be accelerated with PyTorch's GPU acceleration support. 
(4) PyTorch's native automatic differentiation can be applied to train parameters. 

Furthermore, \quantumengine supports push-the-button conversion between PyTorch quantum circuit and IBM Qiskit QuantumCircuit, such that we can support convenient end-to-end training-to-deployment flow. 
It contains many ready-to-use templates, e.g., random and strongly-entangled layers. 
Parameter shift is also supported for gradient computations.
All steps in \name are implemented with it. 
The library has great potential to accelerate research in parameterized QC, especially for QML, VQE, \etc

Figure~\ref{fig:quantum_engine} shows the training speed of 10-qubit parameterized quantum circuits containing 100 \rx~and 100 \texttt{CRY} gates vs. PennyLane. Since PennyLane processes batch with the 'For' loop, the training speed reduces linearly with the batch size. \quantumengine supports tensorized batch processing on CPU/GPU, so the speed is not severely influenced. 
The training speed is $246$ to 10$^4$ times faster than PennyLane.

%% file: figtex/fig_quantum_engine.tex
\begin{figure}[t]
    \centering
    \includegraphics[width=\columnwidth]{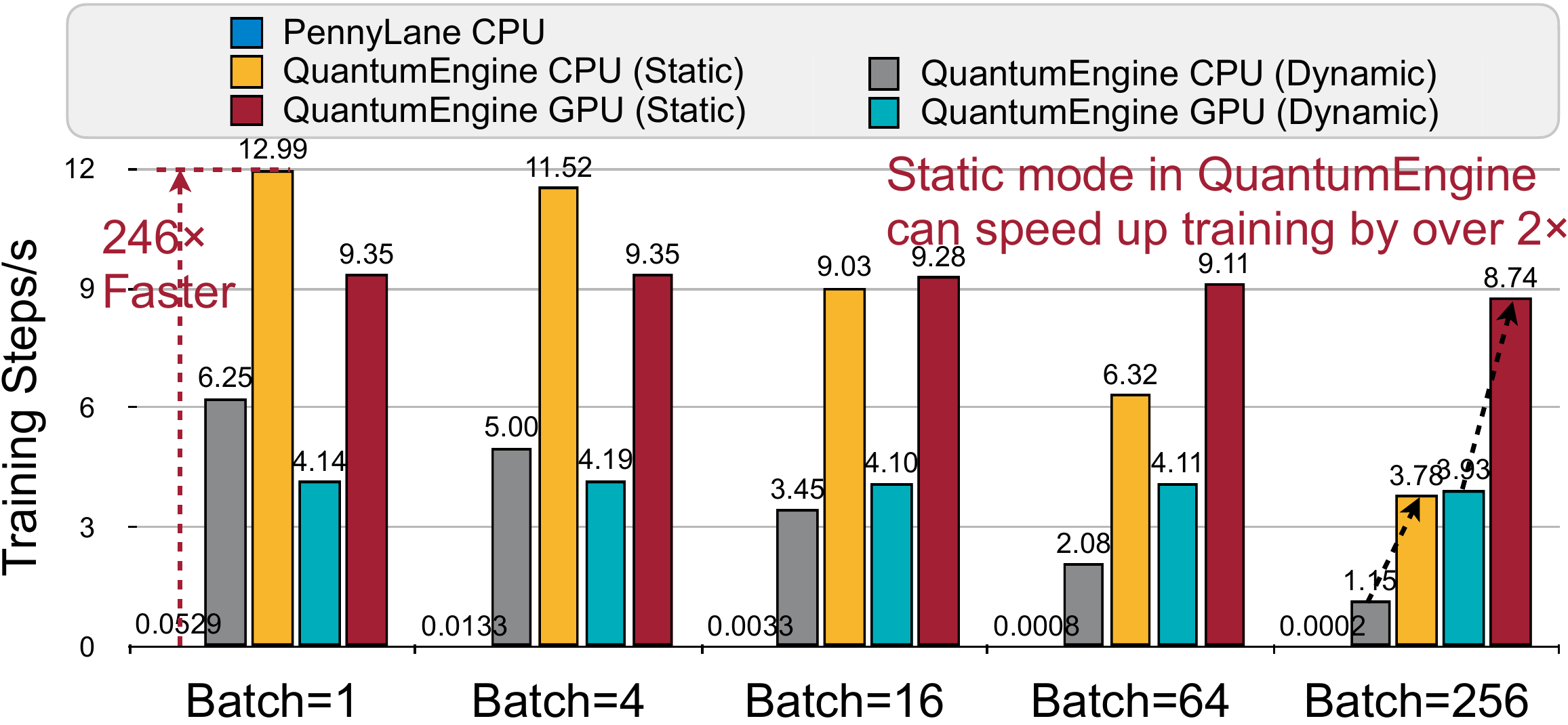}
    \vspace{-20pt}
    \caption{Training speed of \quantumengine \vs PennyLane\cite{bergholm2018pennylane}. \quantumengine provides at least 246$\times$ speedup. 
    }
    \label{fig:quantum_engine}
    \vspace{-15pt}
\end{figure}

%% file: texts/6_evaluation.tex
\section{Evaluation}
\label{sec:evaluation}

\subsection{Evaluation Methodology}

\input{tables/tab_benchmarks}
\textbf{Benchmarks.} We conduct experiments on 6 QML and 6 VQE tasks. QML benchmark information is shown in Table~\ref{tab:benchmarks}.
MNIST and Fashion use 95\% images in `train' split as the training set and 5\% as the validation set. Due to the limited real QC resources, we randomly sample 300 images from the `test' split as our test set and report their accuracy on the real quantum devices. However, we find 300 images can already have comparable accuracy to the whole testing set: on four circuits, the whole testing set acc/300-sample acc are 0.505/0.497, 0.284/0.283, 0.564/0.547, 0.272/0.287. The input images are $28\times28$. We center-crop them to $24\times24$ and down-sample them with average pooling.
Vowel-4 dataset (990 samples) is separated to train:validation:test = 6:1:3 and test with the whole test set. We perform principal component analysis (PCA) for the vowel features and take 10 most significant dimensions. 
For readout, we measure the expectation values on Pauli-\texttt{Z} basis and obtain a value [-1, 1] from each qubit. 
For 2-class, we sum the qubit 0 and 1, 2 and 3 respectively to get two values, which will be processed by Softmax to get probabilities. 
For 4 and 10-class, we use Softmax on expectation values to obtain probabilities.

For VQE, the goal is to find the low-energy eigenvalue of a target molecule by repeated measurements of the expectation value of the Hamiltonian of the molecule (as detailed in Section~\ref{sec:related}). The molecules we study in this work contain \htwo, \lih, \water, \chfour, and BeH$_2$ as in Table~\ref{tab:benchmarks}. 
VQE circuits are searched and trained on classical machines then deployed on real QC to obtain the eigenvalues.

\textbf{Quantum Devices and Compiler Configurations.} We use IBMQ quantum computers via Qiskit~\cite{ibmq} APIs. We study 10 devices, with \#qubits from 5 to 65 and Quantum Volume from 8 to 128. We also employ Qiskit for compilation. The optimization level is set to 2 except for level 3 for Noise-adaptive and Sabre baselines in Figure~\ref{fig:main_acc} and Table~\ref{tab:estimator_qc}. For searched qubit mapping, we set it as the `initial\_layout' of the compiler. QML/VQE experiments run 8192/2048 shots.

\input{tables/tab_circuit_depth}
\input{figtex/fig_main_acc}

\textbf{Circuit Design Spaces.}
We select 6 circuit design spaces, 4 from previous QML work, and name them with gates: 
\begin{enumerate}[leftmargin=*]
\setlength\itemsep{-0.2em}
    \item `\uthree+\cuthree' -- One block has a \uthree~layer with one \uthree~gate on each qubit, and a \cuthree~layer with ring connections, \eg \cuthree(0, 1), \cuthree(1, 2), \cuthree(2, 3), \cuthree(3, 0).
    \item  `\zz+\ry'~\cite{lloyd2020quantum} -- One block contains one layer of \zz~gate, also with ring connections, and one \ry~layer. 
    \item  `\rxyz'~\cite{mcclean2018barren} -- One block has four layers: \rx, \ry, \rz, and \cz. There is one \sqrth~layer before the blocks. 
    \item  `\zx+\xx'~\cite{farhi2018classification} -- according to their MNIST circuit design, one block has two layers: \zx~and \xx. 
    \item  `\rxyz+\uone+\cuthree'~\cite{henderson2020quanvolutional} -- according to their random circuit basis gate set, we design \supercircuit in which one block has 11 layers in the order of \rx, \sgate, \cnot, \ry, \tgate, \swap, \rz, \hgate, \sqrtswap, \uone~and \cuthree.
    \item `\texttt{IBMQ Basis}'~\cite{ibm_2021} -- we design SuperCircuit with basis gate set of IBMQ devices, in which one block has 6 layers in the order of \texttt{RZ}, \texttt{X}, \texttt{RZ}, \texttt{SX}, \texttt{RZ}, \texttt{CNOT}.
\end{enumerate}
The \supercircuits for space 1 to 4 contain 8 blocks; space 5 has 4 blocks; space 6 has 20 and does not have front sampling. The design spaces contain numerous \subcircuits, e.g.: \rxyz+\uone+\cuthree~contains $4^{11\times4}=3\times10^{26}$ \subcircuits.

\input{figtex/fig_devices}
\textbf{Baselines.}
We have six baselines: (1) \emph{Noise-unaware search}: the \subcircuits are searched with noise-free simulation. No noise information is involved. (2) \emph{Random generation}: with the same gate set, we generate random circuits and constrain their \#parameters the \emph{same} as the \name searched circuit for fair comparisons. We generate three different circuits and report the best. (3) \emph{Human design}: we also make sure the same \#parameters. For \uthree+\cuthree, \rxyz+\uone+\cuthree~and \texttt{IBMQ Basis} spaces, human design has full width in the several front blocks. For \zz+\ry, \rxyz, and \zx+\xx~spaces, we stack multiple blocks introduced in the original paper. 
The last layer of human designs may not have full width to make sure the same total number of parameters.
(4) \emph{Human design+noise-adaptive mapping}: the circuit has the same \#parameters with QuantumNAS. The qubit mapping is optimized with state-of-the-art technique~\cite{murali2019noise}. (5) \emph{Human design+Sabre mapping}: the circuit has the same \#parameters with QuantumNAS, the qubit mapping is optimized with Sabre~\cite{li2019tackling}. (6) \emph{Human design(1/2 \#Param)+Sabre mapping}: similar to (6) with half \#parameters. (7) For VQE, we have an additional \emph{UCCSD}~\cite{bartlett2007coupled} baseline. For UCCSD of CH$_4$-10Q and BeH$_2$, the original circuit cannot be successfully run on IBMQ machines because of too many gates ($>$10,000), so we only take the front 1,000 gates for real QC experiments and report the results of the full circuit using the Qiskit noisy simulator.

\textbf{\supercircuit and \subcircuit Training Setups.}
For all searched \subcircuits and baselines, we use the same training setting for fair comparisons. We use Adam optimizer with initial learning rate 5e-3 and weight decay 1e-4, cosine learning rate scheduler. We train for 200 epochs with batch size 256 for QML tasks; 1000 steps for VQE tasks with batch size 1. For QML, the objective is to minimize training loss, while VQE minimizes the eigenvalue. 
\supercircuits training has the same settings with \subcircuits, except adding a linear learning rate warm-up from 0 to 5e-3 in the first 30 epochs for QML and 150 steps for VQE. Restricted sampling is applied during the whole training process. We set the largest number of different layers as seven. An additional technique is to progressively shrink the lower bound of possible sampled \subcircuit \#blocks to stabilize training. We use Nvidia TITAN RTX 2080 GPU. The time cost is shown in Figure~\ref{fig:overview}.

\input{tables/tab_hardware_specific}

\textbf{Noise-Adaptive Evolutionary Co-Search Setups.}
The evolutionary search is conducted with inherited gate parameters on the validation set of QML tasks. For QML and VQE, the evolution engine searches 40 iterations with a population of 40, parents population 10, mutation population 20 with 0.4 mutation probability, and crossover population 10. 
The noise model is obtained from IBM's calibration data for the performance estimator, and the noise simulator is the Qiskit QASM simulator. We also run 8192 shots on simulators.

\textbf{Iterative Pruning Setups.}
The searched \subcircuit is firstly trained from scratch. In pruning, we set five final pruning ratios, 0.1, 0.2, 0.3, 0.4, and 0.5. The starting ratio is 0.05. Pruning starts at step 0 and ends at half of total steps. We report the highest measured accuracy among the five ratios.

\subsection{Experimental Results}

\textbf{Results on Four and Two Classifications.}
Figure~\ref{fig:main_acc} shows the measured accuracy on IBMQ-Yorktown (5Q) of \name and 6 baselines on 5 QML tasks in 6 different design spaces. \name achieves over 85\% 4-class and 95\% 2-class accuracy and consistently outperforms baselines except for Vowel-4 in \zz+\ry~space and MNIST-4 in \zx + \xx~space. 
The statistics for Fashion-2 \uthree+\cuthree~space are in Table~\ref{tab:depth}. The noise-unaware search only optimizes noise-free accuracy, which results in a deep circuit (237 depth) with low measured accuracy. 
\uthree+\cuthree, \rxyz, \rxyz+\uone+\uthree~and \texttt{IBM Basis} are better spaces as they always outperform the remaining two design spaces, and thus they are considered more noise-resilient. 
In addition, pruning brings an average of 2\% for 4-class and 1\% improvement for 2-class tasks. When the searched circuits contain only a small number of parameters, such as 7 in Vowel-4 \zz+\ry, removing any parameter will hurt the accuracy. For circuits with more parameters, such as 36 for MNIST-4 \uthree+\cuthree, the pruning ratio can be 50\% while increasing accuracy by 4\%. In Table~\ref{tab:depth}, pruning removes 14 parameters and reduces depth by 11. The accuracy is improved by 3\% since the pruned circuit has similar noise-free accuracy but fewer gates and less noise. For \texttt{IBMQ Basis}, although its space is larger than \texttt{U3+CU3}, the accuracy is sometimes lower. Hence, a larger design space does not necessarily bring better final performance because of the higher search difficulty.
\input{figtex/fig_large_devices}
\input{figtex/fig_mnist_train_curve}

\textbf{Results on Different Quantum Devices and Noise.} Figure~\ref{fig:devices} shows QuantumNAS performance on various devices. 
For one task, QuantumNAS SubCircuits for each device are searched with the same SuperCircuit, but with noise models tailored for each device.
For the machine with the smallest noise, IBMQ-Santiago, although the baseline methods achieve higher accuracy than Melbourne and Guadalupe, QuantumNAS can still deliver 5\% better accuracy on average. Additionally, we show the accuracy of \name tested 3 weeks after search, which is slightly lower than tested immediately but still much higher than baselines. One reason is that the machines are calibrated by the IBMQ at least once a day, so noises are not far from the calibration point. Therefore, even the noise characteristics change on a machine in calibration interval, \name circuits are still noise-resilient.
The results on Athens are unavailable since it is retired.  Table~\ref{tab:hardware_specific} shows performance of circuits searched and run on different devices. Best performance is achieved when two devices are the same, which shows the necessity of device-specific circuits.

\textbf{Scalability.} 
We further show \name results on larger machines with larger circuits. 
We search for circuits with 15, 16, 21, 21 qubits in \texttt{U3+CU3} space for machines with 15, 16, 27, 65 qubits in Figure~\ref{fig:large_devices}. For the 21 qubit model, the SuperCircuit contains 1 block. \name can achieve over 5\% better accuracy. For even larger circuits for which classical simulations are infeasible, we can move the \emph{whole pipeline} to quantum machines. Super/Subcircuit training can be done with parameter shift, and evolutionary search can directly evaluate \subcircuits on quantum machines. We demonstrate the high feasibility of training circuits on quantum machines using parameter shift. We train SubCircuits for different tasks on different machines as in Table~\ref{tab:paramshift}. Results show comparable real QC test accuracy of training on real QC and classical simulators. 
We also show the training curve of MNIST-4 in Figure~\ref{fig:scalability} left.

We add experiments on using real QC devices to evaluate SubCircuits in search as in Table~\ref{tab:estimator_qc} and compare with using noisy simulator. We experiment with Qiskit optimization levels 2 and 3. 
Due to queuing, we can only afford 20 search iterations which take $\sim$3 days. 
The accuracy of using real QC is similar to using simulators. In addition, the opt. level 3 cannot consistently improve accuracy over level 2. The observation aligns with recent work on QC characterization~\cite{10.5555/3433701.3433762}. One reason is that \name has already found a good mapping that is hard to optimize further. 
We further show the runtime of QuantumNAS training a VQE model fully on real QC with different qubit numbers in Figure~\ref{fig:scalability} right. The SuperCircuit contains 32 parameters in \texttt{ZZ+RY} space. The runtime is an \emph{upper bound} as we assume the largest SubCircuit is used. The pruning ratio is set as 50\%. The runtime increases approximately \textbf{linearly} as the qubit number increases. The projected 127 qubit runtime is around 57 hours. Thus QuantumNAS has \textbf{high scalability.}

\input{tables/tab_estimator_qc}
\input{tables/tab_paramshift}

\textbf{Results on VQE Tasks.}
Figure~\ref{fig:vqe_h2} shows the VQE performance for \htwo in different spaces, measured on the IBMQ-Yorktown. The theoretical optimal value is -1.85. Estimated eigenvalues obtained by \name are \emph{consistently lower} than any other baselines. The UCCSD ansatz baseline is far from the optimal value as it is not adapted to the hardware noises. Pruning removes 50\% parameters for all five circuit design spaces and can steadily reduce eigenvalues. Thus VQE circuits have a higher degree of redundancy over QML ones, making them \emph{more amenable to pruning}.
Figure~\ref{fig:vqe_lih} further shows the comparison results of \name and UCCSD on \lih(6Q), \water(6Q), CH$_{4}$(4Q and 10Q) and BeH$_{2}$(15Q) on machines with 7Q, 15Q, and 27Q. Besides achieving lower measured expectation values, \name can also reduce the theoretically trained values. For \water, the UCCSD noise-free trained expectation value is -49.6 while \name has -52.4, indicating that \name ansatz adapts to both the device and the molecule -- a \emph{hybrid device and problem ansatz}. For CH$_{4}$-10Q, we also use the IBMQ-Montreal noisy simulator to simulate the full original circuit (7164 gates) and obtains expectation value as -12.86; We also get that of BeH$_{2}$-15Q (30851 gates) as -9.81. Despite the much larger circuit, the full circuit results are \textbf{worse} than shallower ones because the larger number of gates introduce more significant noise.

\input{figtex/fig_vqe_h2}

\input{figtex/fig_vqe_lih}

\input{figtex/fig_codesign_effect}
\input{figtex/fig_progressive_effect}

\input{figtex/fig_topoerror}
\subsection{Performance Analysis}
\label{subsec:perf}

\textbf{Accuracy Improvement Breakdown.}
We select five tasks and five design spaces to show the breakdown of accuracy improvements in Figure~\ref{fig:codesign_effect}. We compare the \name co-search to three baselines: (1) human baseline with no circuit or qubit mapping search, (2) noise-adaptive mapping search only, and (3) noise-adaptive circuit search only. Only searching circuit has larger accuracy improvements than only searching qubit mapping, as the space for circuit search is much larger, echoing our motivation. \emph{The co-design of both aspects can further unlock 9\% accuracy gain on average.} 
As already mentioned in Section~\ref{sec:introduction}, Figure~\ref{fig:curve} further shows the \#parameters \vs accuracy curves. 
With more parameters, the accuracy of baseline designs quickly saturates and drops due to gate errors. In contrast, \name can mitigate the negative effect of gate errors, and \emph{delays the accuracy peak}, enabling more effective circuit parameters and higher accuracy.

\textbf{Effect of Front and Restricted Sampling.}
Figure~\ref{fig:progressive_effect} shows the measured performance of \subcircuits on five tasks in \zx+\xx~and \rxyz+\uone+\cuthree~spaces. The baseline method is random sampling. Since the front and restricted sampling controls the difference between the consecutive samples, the SuperCircuit training is more stable. Thus it improves the \emph{reliability} of estimated relative performance, the searched \subcircuit is closer to the optimal one and achieves on average 12\% higher final accuracy.

\input{figtex/fig_evo_scatter}

\input{figtex/fig_pruning_curve}
\input{tables/tab_small_space}
\input{tables/tab_prune}

\textbf{Effect of qubit topology/error rate/qubit mapping to performance and design choice.} Figure~\ref{fig:topoerror} shows MNIST-4 and $H_{2}$ VQE performance on devices with various topologies and error rates. We have observations: (1) Comparing Santiago, Rome, and Athens, with the same topology, a lower error rate brings better performance. Yorktown has the highest error rate so the performance is worse than others. (2) Comparing Rome (`--') and Lima (`T'), Quito (`T') and Yorktown(`+'), under similar error rates, `T' topology brings better performance than the other two. (3) For qubit choices (mapping), the co-searched mapping can consistently outperform the na\"ive mapping. (4) For design choices of co-search, the average convergence iteration is 13.5, 14, 9.2 for `T', `+', and `--' respectively. Therefore, we need a relatively larger iteration number for co-search on topology `T' and `+' machines. That may be due to their more complicated connections than `--'.

\textbf{Search in Small Design Space.} We construct a small \texttt{U3+CU3} space that does not break into multiple blocks. All SubCircuits can be arbitrarily sampled without front sampling. The circuit depth is around 40. Comparisons with larger space with multiple blocks are shown in Table~\ref{tab:small_space}. Small space has consistently worse accuracy: although small circuits have less noise, it also has \emph{smaller learning capacity}. QuantumNAS can find a better trade-off between noise and capacity. This can only be achieved when QuantumNAS has access to a relatively large design space. So we cannot only search shallow circuits.

\textbf{Random Search \vs Evolutionary Search.} 
Multiple candidate algorithms are applicable for the search stage. We compare the evolutionary with random search in Figure~\ref{fig:evo_scatter}. The best performance of random search quickly saturates, while evolutionary can find \subcircuit and qubit mapping pair with lower loss, which delivers higher accuracy.

\textbf{Effect of Pruning Ratios.}
Figure~\ref{fig:pruning_curve} shows the effect of different final pruning ratio for MNIST-2 \zz+\ry~space and Fashion-2 \uthree+\cuthree~spaces. As the final ratio increases, there exists a sweet spot where the positive effect of gate error reduction can overcome the negative effect of smaller circuit capacity so we can observe a peak accuracy. In general, circuits with more parameters can afford a larger pruning ratio; those with fewer parameters has a smaller ratio. 
In Table~\ref{tab:prune}, we further show the acceleration from pruning a circuit with 180 parameters using the Pennylane classical simulation framework.

%% file: tables/tab_benchmarks.tex
\begin{table}[t]
\centering

\caption{Benchmark information summary.}
\vspace{-10pt}
\label{tab:benchmarks}
\resizebox{\columnwidth}{!}{%
\begin{tabular}{lcccc}
\toprule
Task & Class & Input Size & \#qubit & Encoder (Mapping) \\
\midrule
MNIST-10 & \texttt{0-9} & 6$\times$6 & 10 & 10\ry,10\rz,10\rx,6\ry \\
MNIST-4 & \texttt{0,1,2,3} & 4$\times$4 & 4 & 4\ry,4\rz,4\rx,4\ry \\
MNIST-2 & \texttt{3,6} & 4$\times$4 & 4 & 4\ry,4\rz,4\rx,4\ry \\
\multirow{2}{*}{Fashion-4} & \texttt{t-shirt,trouser} & \multirow{2}{*}{4$\times$4} & \multirow{2}{*}{4} & \multirow{2}{*}{4\ry,4\rz,4\rx,4\ry} \\
& \texttt{pullover,dress} & & &\\
Fashion-2 & \texttt{dress,shirt} & 4$\times$4 & 4 & 4\ry,4\rz,4\rx,4\ry \\
Vowel-4 & \texttt{hid,hId,had,hOd} & 10 & 4 & 4\ry,4\rz,2\rx \\ 
\midrule
\htwo & VQE & -- & 2 & Bravyi-Kitaev~\cite{bravyi2002fermionic} \\
\water & VQE & -- & 6 & Bravyi-Kitaev \\
\lih & VQE & -- & 6 & Bravyi-Kitaev \\
\chfour-6Q & VQE & -- & 6 & Bravyi-Kitaev \\
\chfour-10Q & VQE & -- & 10 & Bravyi-Kitaev \\
\behtwo & VQE & -- & 15 & Bravyi-Kitaev \\
\bottomrule
\end{tabular}%
}
\vspace{-15pt}
\end{table}

%% file: tables/tab_circuit_depth.tex
\begin{table}[t]
\centering
\renewcommand*{\arraystretch}{1}
\setlength{\tabcolsep}{7pt}
\footnotesize
\caption{Compiled circuit properties for \name and baselines.
}
\vspace{-8pt}

\begin{tabular}{lcccc}
\toprule
& Depth & \#Gates (\#1Q+\#CNOT) & \#Params & Acc. \\
\midrule
Noise-Unaware  & 237 & 365 (299+66) & 120 & 0.48 \\ 
Random & 45 & 100 (94+6) & 36 & 0.86 \\
Human & 64 & 135 (124+11) & 36 &  0.88 \\
\textbf{\name} & 70 & 133 (123+10) & 36 & \textbf{0.89} \\
\textbf{\ \ \ + Pruning} & 59 & 116 (106+10) & 22 & \textbf{0.92} \\
\bottomrule

\end{tabular}

\label{tab:depth}
\vspace{-15pt}

\end{table}

%% file: figtex/fig_main_acc.tex
\begin{figure*}[t]
    \centering
    \includegraphics[width=\textwidth]{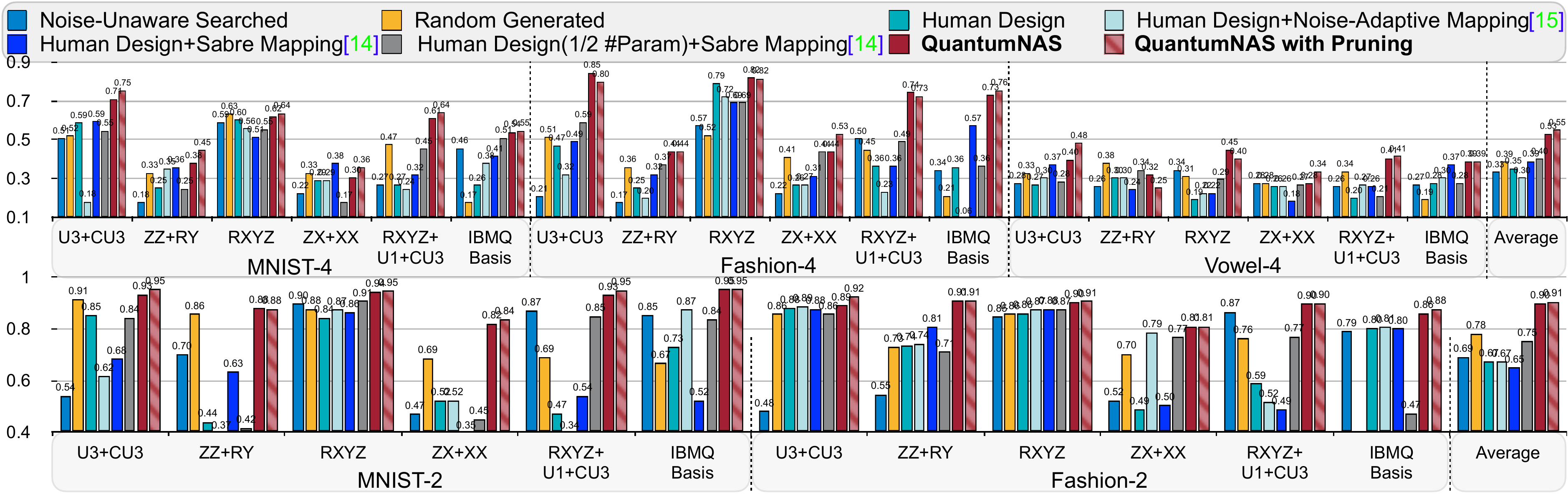}
    \vspace{-20pt}
    \caption{QuantumNAS achieves the highest accuracy on real QC devices (IBMQ-Yorktown). Pruning further improves 2\% on average.}
    \label{fig:main_acc}
    \vspace{-10pt}
\end{figure*}

%% file: figtex/fig_devices.tex
\begin{figure*}[t]
    \centering
    \includegraphics[width=\textwidth]{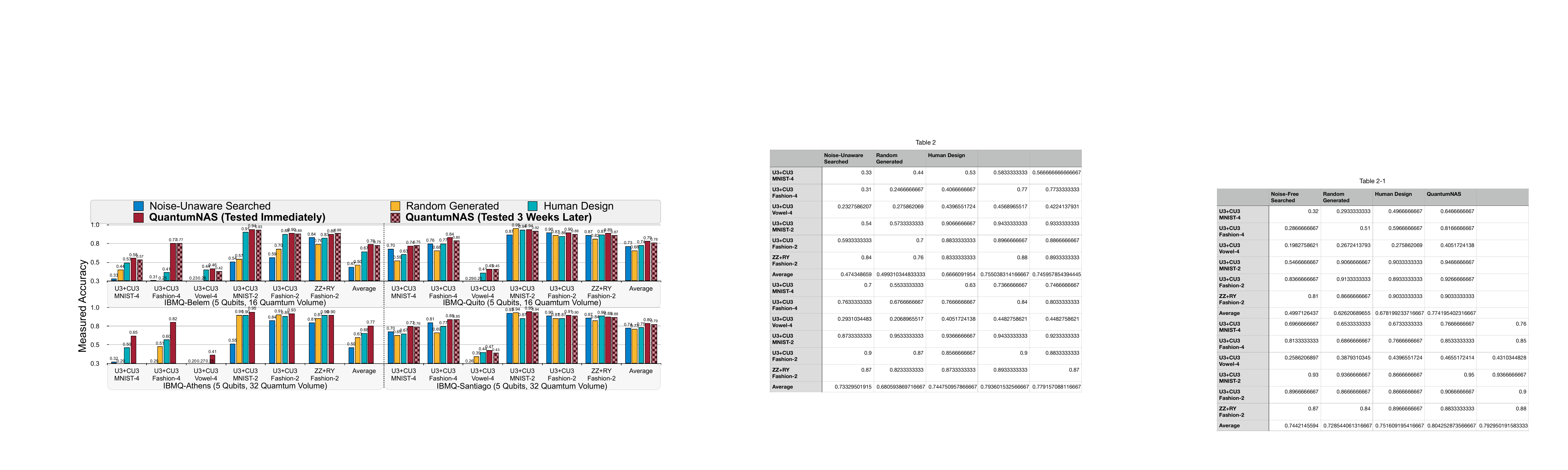}
    \vspace{-20pt}
    \caption{
    On four 5-Qubit real QC devices, the design searched by QuantumNAS outperforms baseline designs with higher measured accuracy.
}
    \label{fig:devices}
    \vspace{-10pt}
\end{figure*}

%% file: tables/tab_hardware_specific.tex
\begin{table}[t]

\centering
\renewcommand*{\arraystretch}{1}
\setlength{\tabcolsep}{7pt}
\footnotesize

\caption{Device-specific circuit has the best accuracy.}
\vspace{-8pt}
\begin{tabular}{lcccc}
\toprule
\small \underline{Run on $\downarrow$} Searched for $\rightarrow$ &  \multirow{1}{*}{Yorktown}  & \multirow{1}{*}{Belem} & \multirow{1}{*}{Santiago}  \\
        \midrule
        
        Yorktown & \cellcolor{blue!20} \textbf{0.85} & \cellcolor{myred!20} 0.60  & 
        \cellcolor{myred!20} 0.54 \\
        Belem & \cellcolor{myred!20} 0.67  &  \cellcolor{blue!20} \textbf{0.77} & \cellcolor{myred!20} 0.43 \\
        
        Santiago &\cellcolor{myred!20} 0.82 & \cellcolor{myred!20} 0.81 & \cellcolor{blue!20} \textbf{0.85} \\                        \bottomrule
        
\end{tabular}%

\vspace{-15pt}
\label{tab:hardware_specific}


\end{table}

%% file: figtex/fig_large_devices.tex
\begin{figure*}[t]
    \centering
    \includegraphics[width=\textwidth]{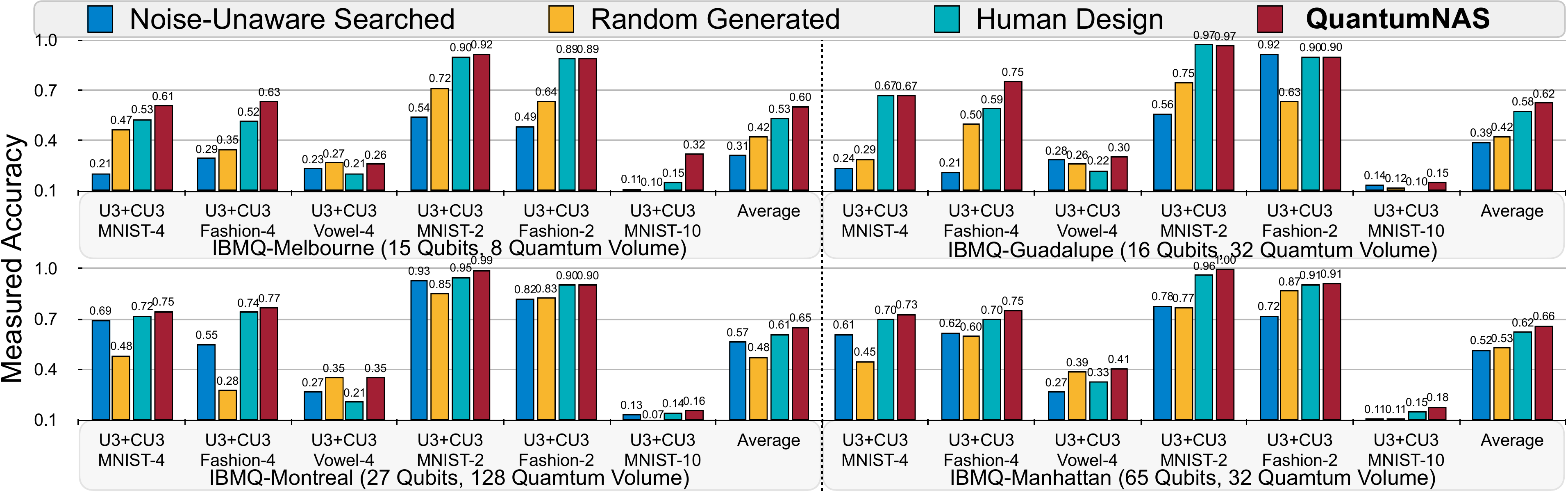}
    \vspace{-20pt}
    \caption{
    QuantumNAS has high scalability and can maintain the performance advantage with large models on large devices (in terms of \#qubits).
}
    \label{fig:large_devices}
    \vspace{-15pt}
\end{figure*}

%% file: figtex/fig_mnist_train_curve.tex
\begin{figure}[t]
    \includegraphics[width=\columnwidth]{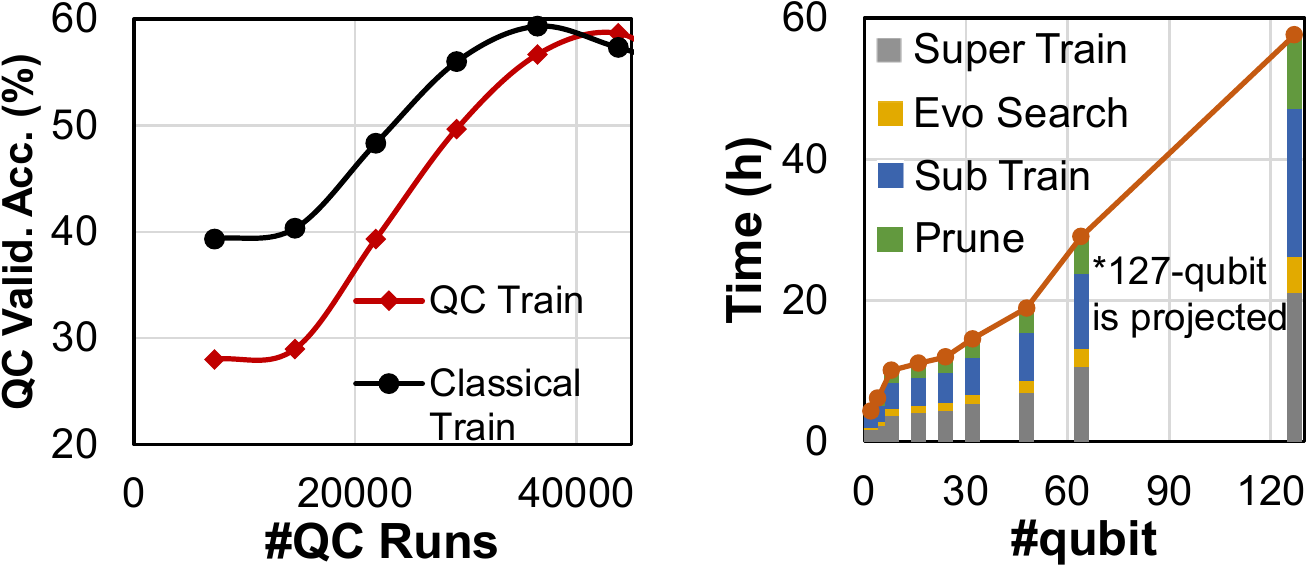}
    \vspace{-20pt}
    \caption{
    (a) Scalable QNN training for MNIST-4 on real quantum machines with parameter shift.
    (b) Runtime breakdown with different numbers of qubits.}
    \label{fig:scalability}
    \vspace{-10pt}
\end{figure}

%% file: tables/tab_estimator_qc.tex
\begin{table}[t]
\centering

\centering
\renewcommand*{\arraystretch}{1}
\setlength{\tabcolsep}{3pt}
\footnotesize
\caption{Search with estimators vs. real QC for Fashion-4 \texttt{U3+CU3}.}
\vspace{-8pt}
\resizebox{\columnwidth}{!}{%
\begin{tabular}{lcccccccccc}
\toprule
 \multirow{2}{*}{Method}   & \multicolumn{5}{c}{Optimization Level 2}                                              & \multicolumn{5}{c}{Optimization Level 3}                              \\ \cmidrule(lr){2-6}\cmidrule(lr){7-11} 
                          & York.      & Bel.         & Qui.         & Ath.        & Sant.      & York.      & Bel.     & Qui.     & Ath.    & Sant.  \\ \midrule
\multicolumn{1}{c}{Est.}  & \textbf{0.85} & 0.77          & \textbf{0.84} & \textbf{0.82} & \textbf{0.77} & 0.69 & \textbf{0.63}          &      \textbf{0.82}     & 0.84          & \textbf{0.86}          \\ \midrule
\multicolumn{1}{c}{Real QC} & 0.66          & \textbf{0.80} & 0.76          & 0.77          & 0.73          &         \textbf{0.70}      & 0.54& 0.72 & 0.84 & 0.85 \\ \bottomrule
\end{tabular}%
}

\vspace{-10pt}
\label{tab:estimator_qc}
\end{table}

%% file: tables/tab_paramshift.tex
\begin{table}[t]
\centering
\renewcommand*{\arraystretch}{1}
\setlength{\tabcolsep}{1pt}

\caption{Circuit training on real QC with parameter shift is feasible.}
\label{tab:paramshift}
\resizebox{\columnwidth}{!}{%
\begin{tabular}{lccccccc}
\toprule
Task & MNIST-4 & MNIST-2 & Fashion-4 & Fashion-2 & Vowel-4 & Fashion-2 & Fashion-2 \\
Machine & Jarkata & Jarkata & Manila & Santiago & Lima & Guadalupe & Montreal \\
Qubit Usage & 4 & 4 & 4 & 4 & 4 & 16 & 16 \\
\midrule
Classical & 0.59 & 0.79 & 0.54 & 0.89 & 0.31 & 0.70 & 0.74  \\
QC Train & 0.59 & 0.83 & 0.49 & 0.84 & 0.34 & 0.71 & 0.74 \\
\bottomrule
\end{tabular}%
}
\end{table}

%% file: figtex/fig_vqe_h2.tex
\begin{figure}[t]
    \centering
    \includegraphics[width=\columnwidth]{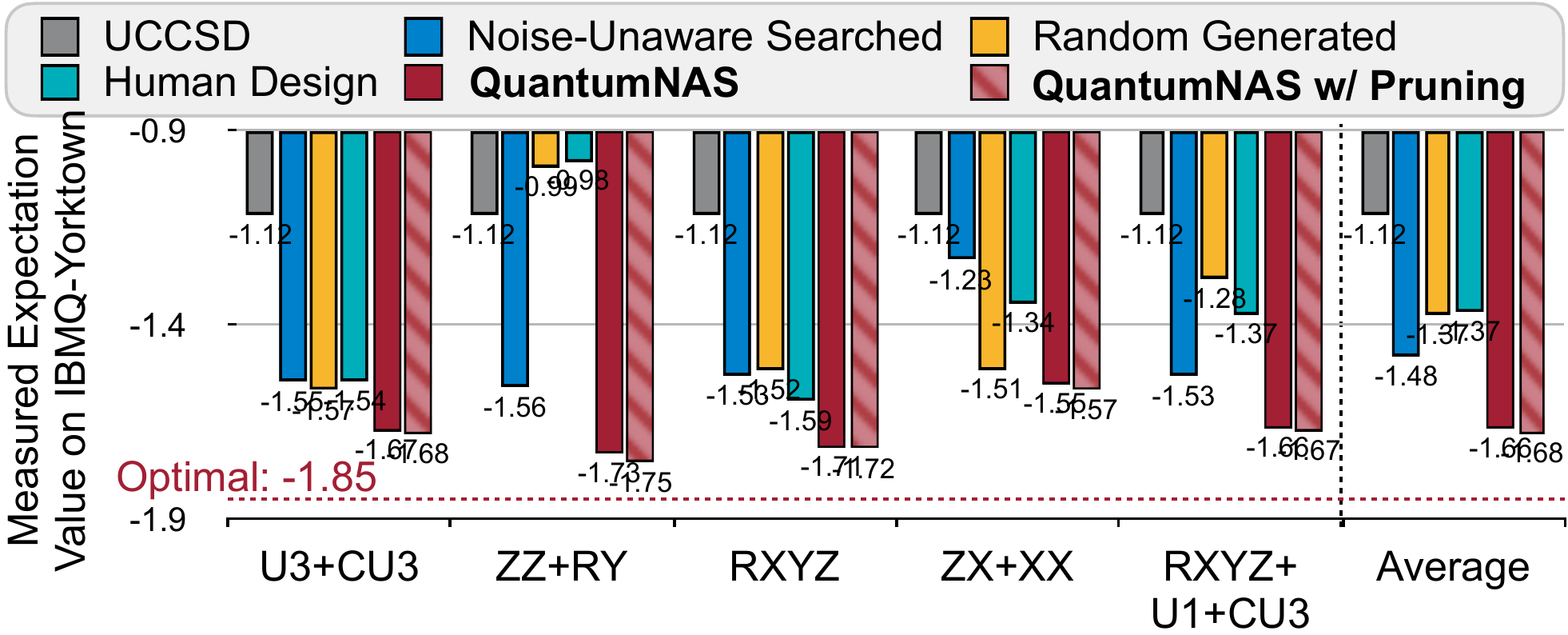}
    \vspace{-20pt}
    \caption{\htwo VQE expectation value in different spaces. 
    \name consistently obtains the lowest expectation value. 
    }
    \label{fig:vqe_h2}
    \vspace{-10pt}
\end{figure}

%% file: figtex/fig_vqe_lih.tex
\begin{figure}[t]
    \centering
    \includegraphics[width=\columnwidth]{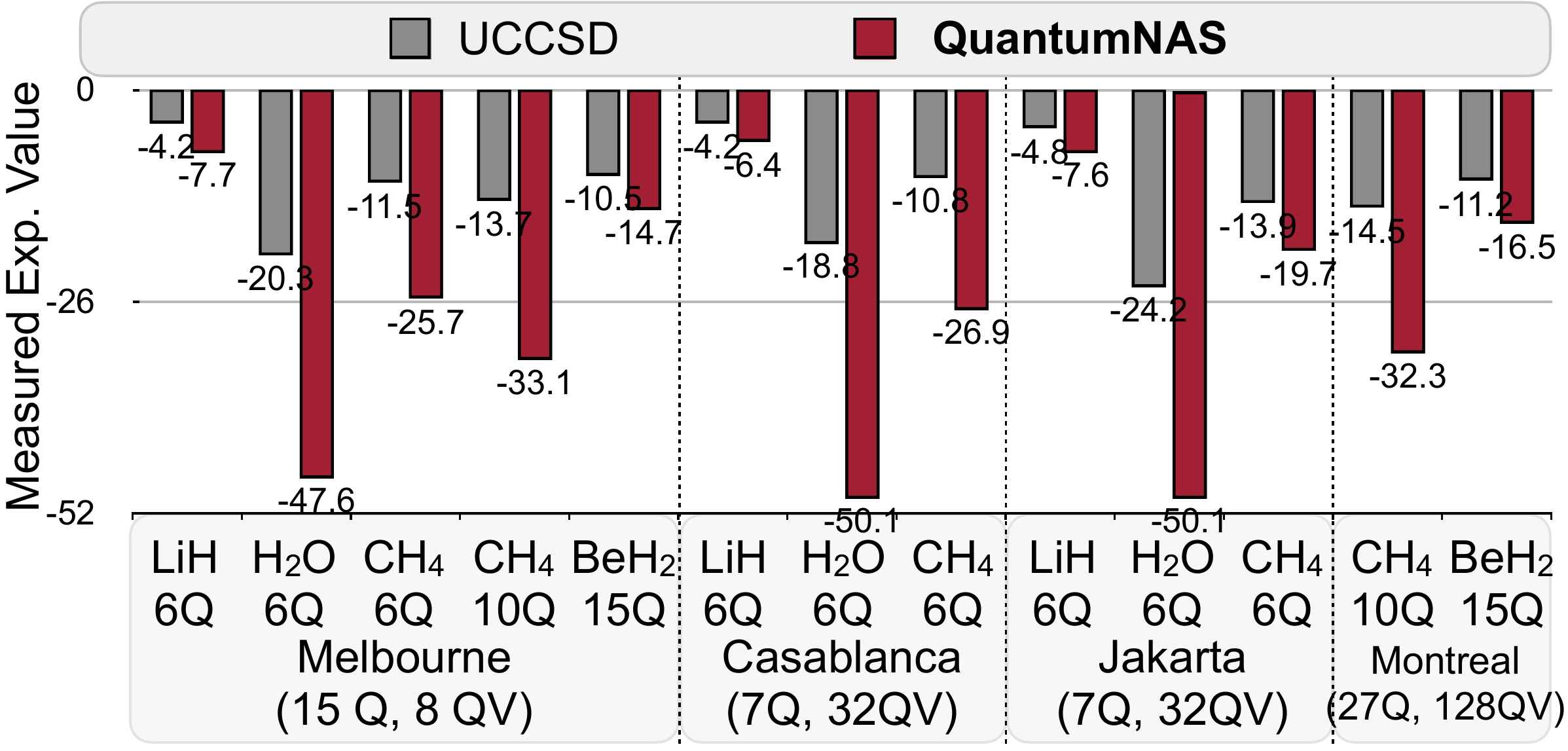}
    \vspace{-20pt}
    \caption{VQE expectation value comparisons on various molecules.
    }
    \label{fig:vqe_lih}
    \vspace{-15pt}
\end{figure}

%% file: figtex/fig_codesign_effect.tex
\begin{figure}[t]
    \centering
    \includegraphics[width=\columnwidth]{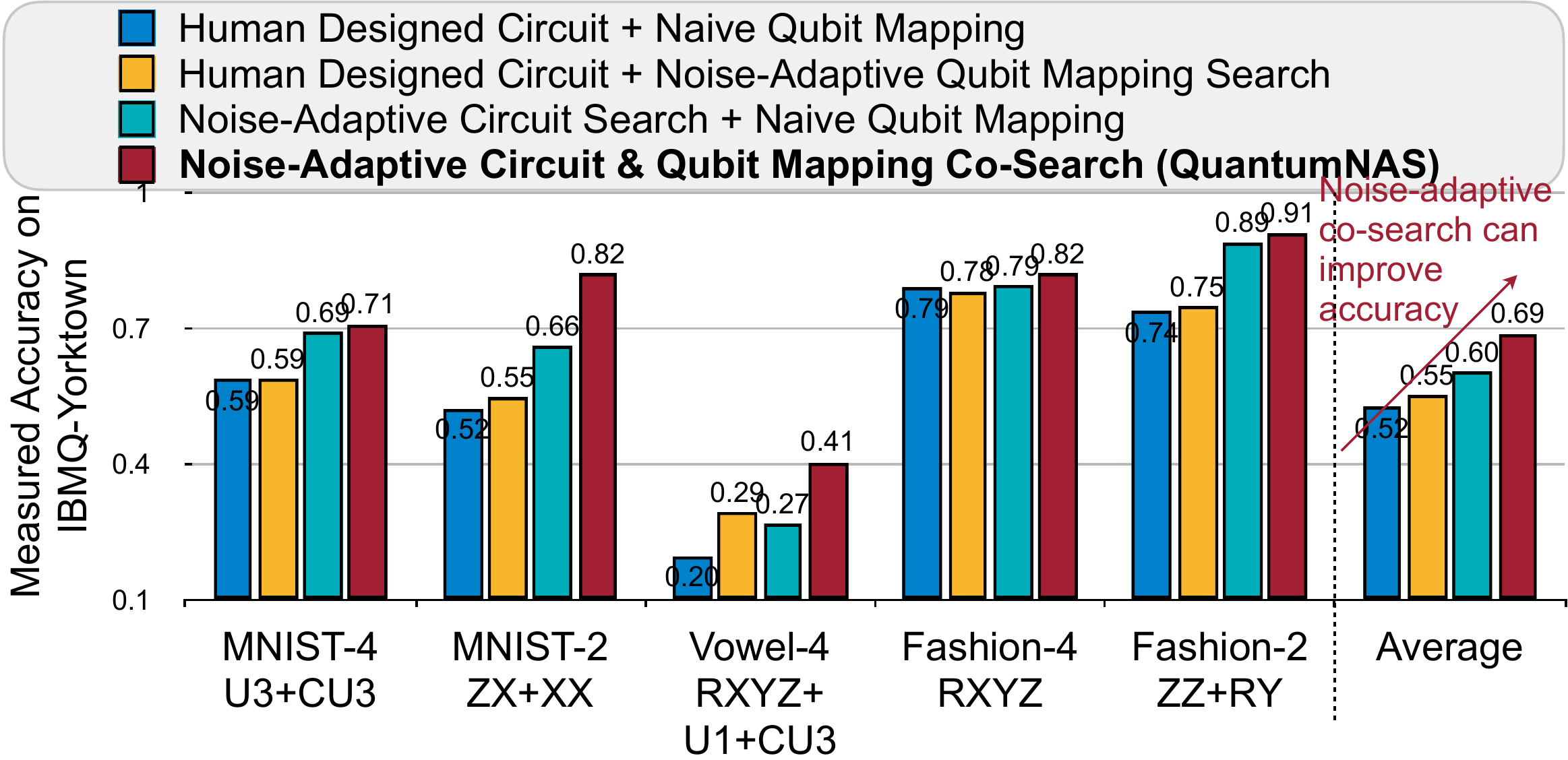}
    \vspace{-20pt}
    \caption{Proposed circuit \& qubit mapping co-search improves accuracy.
    }
    \label{fig:codesign_effect}
    \vspace{-10pt}
\end{figure}

%% file: figtex/fig_progressive_effect.tex
\begin{figure}[t]
    \centering
    \includegraphics[width=\columnwidth]{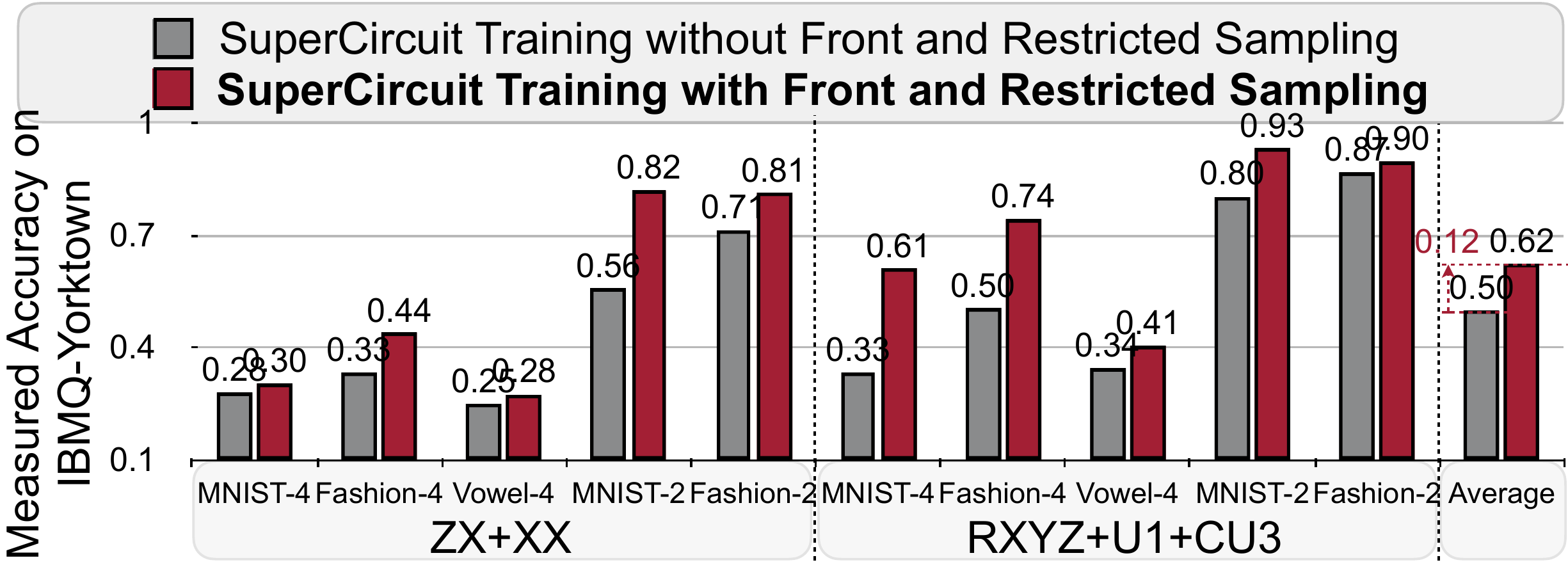}
    \vspace{-20pt}
    \caption{Proposed front and restricted sampling improve final accuracy.
    }
    \label{fig:progressive_effect}
    \vspace{-10pt}
\end{figure}

%% file: figtex/fig_topoerror.tex
\begin{figure}[t]
    \centering
    \includegraphics[width=\columnwidth]{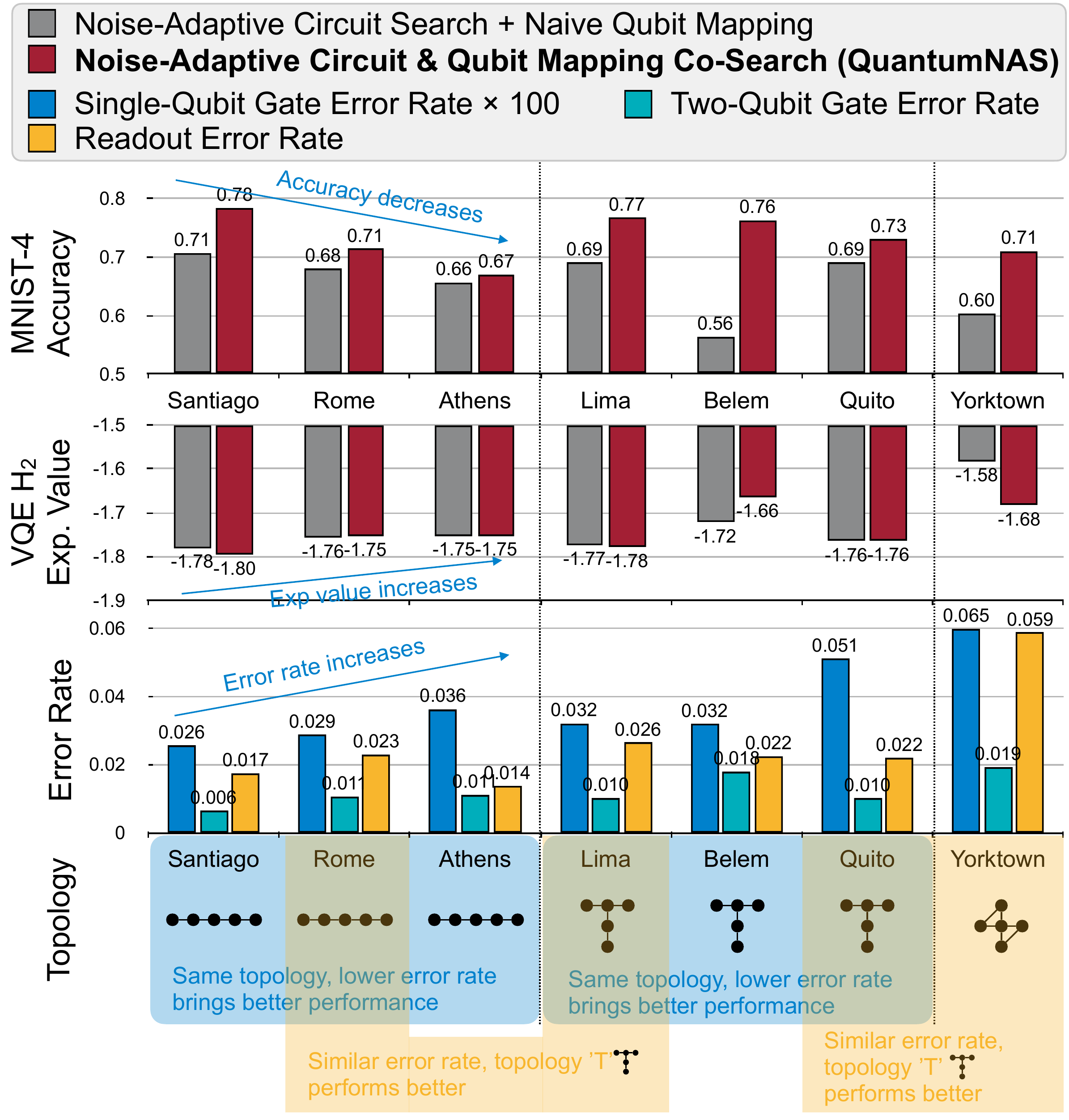}
    \vspace{-20pt}
    \caption{Effect of qubit topology/error rate/qubit mapping on performance.}
    \label{fig:topoerror}
    \vspace{-10pt}
    
\end{figure}

%% file: figtex/fig_evo_scatter.tex
\begin{figure}[t]
    \centering
    \includegraphics[width=\columnwidth]{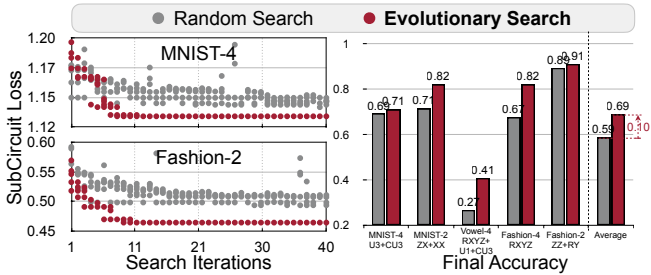}
    \vspace{-20pt}
    \caption{Optimization curves and final accuracy of random/evo search.}
    \label{fig:evo_scatter}
    \vspace{-10pt}
\end{figure}

%% file: figtex/fig_pruning_curve.tex
\begin{figure}[t]
    \centering
    \includegraphics[width=\columnwidth]{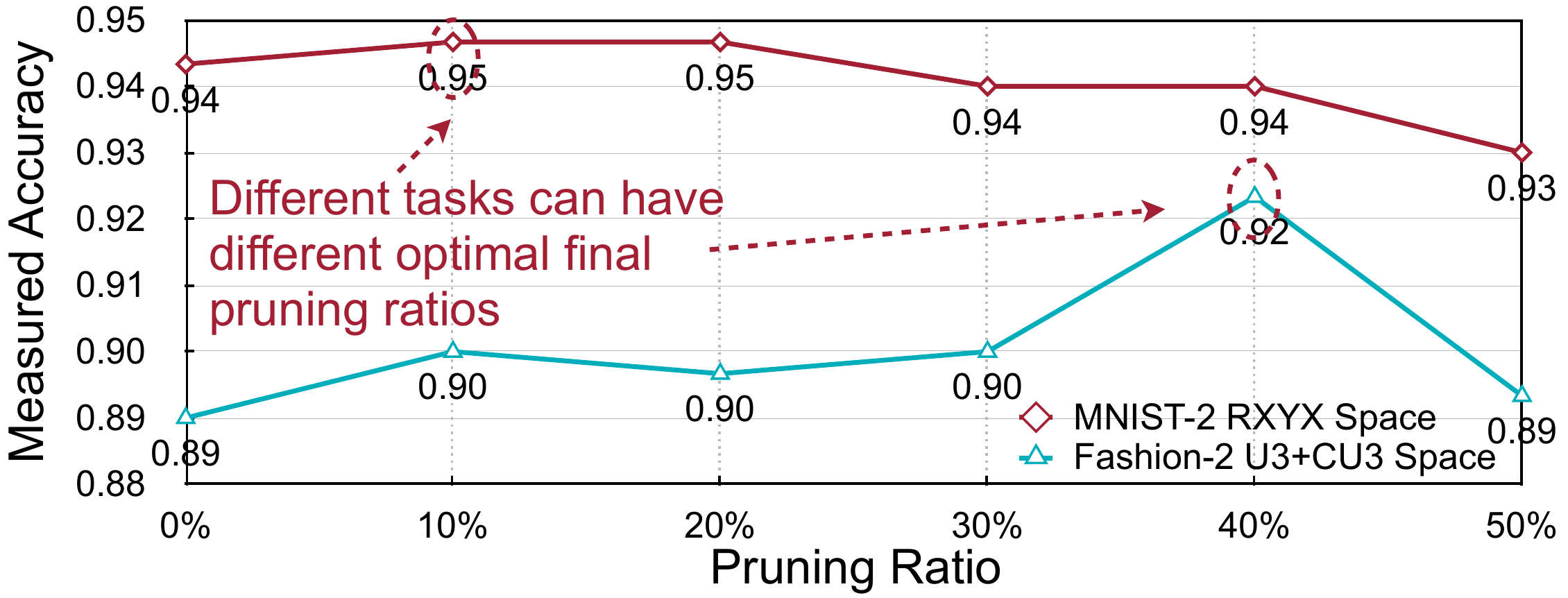}
    \vspace{-20pt}
    \caption{Measured accuracy with different final pruning ratios. 
    }
    \label{fig:pruning_curve}
    \vspace{-10pt}
\end{figure}

%% file: tables/tab_small_space.tex
\begin{table}[t]
\centering
\renewcommand*{\arraystretch}{1}
\setlength{\tabcolsep}{3pt}
\footnotesize
\caption{Ours shows higher acc. than shallow circuits.
}
\vspace{-8pt}
\resizebox{\columnwidth}{!}{%

\begin{tabular}{lccccccccccc}
\toprule
 \multirow{2}{*}{Device}   & \multicolumn{1}{l}{\multirow{2}{*}{Space}} & \multicolumn{2}{c}{\multirow{1}{*}{MNIST-4}} & \multicolumn{2}{l}{\multirow{1}{*}{Fashion-4}} & \multicolumn{2}{c}{\multirow{1}{*}{Vowel-4}} & \multicolumn{2}{c}{\multirow{1}{*}{MNIST-2}} & \multicolumn{2}{c}{\multirow{1}{*}{Fashion-2}} \\

                           & \multicolumn{1}{l}{}                       & \emph{D}               & Acc.                   & \emph{D}                 & Acc.                   & \emph{D}                & Acc.                  & \emph{D}                & Acc                  & \emph{D}                & Acc                    \\ \midrule
\multirow{2}{*}{Santiago}   & Shallow                                       & 50                  & 0.55                  & 58                    & 0.56                  & 35                   & 0.27                 & 28                   & 0.94                 & 30                   & 0.87                   \\  
                           & Ours                                        & \textbf{73}         & \textbf{0.77}         & \textbf{107}          & \textbf{0.85}         & \textbf{116}         & \textbf{0.47}        & \textbf{191}         & \textbf{0.95}        & \textbf{74}          & \textbf{0.91}          \\ \midrule
\multirow{2}{*}{Belem}      & Shallow                                       & 29                  & 0.54                  & 30                    & 0.57                  & 35                   & 0.27                 & 28                   & 0.94                 & 30                   & 0.87                   \\ 
                           & Ours                                        & \textbf{50}         & \textbf{0.58}         & \textbf{68}           & \textbf{0.77}         & \textbf{77}          & \textbf{0.46}        & \textbf{81}          & \textbf{0.94}        & \textbf{62}          & \textbf{0.90}          \\ \midrule
\multirow{2}{*}{Yorktown} & Shallow                                       & 29                  & 0.60                  & 30                    & 0.56                  & 39                   & 0.27                 & 51                   & 0.91                 & 30                   & 0.89                   \\ 
                           & Ours                                        & \textbf{71}         & \textbf{0.71}         & \textbf{82}           & \textbf{0.85}         & \textbf{119}         & \textbf{0.40}        & \textbf{83}          & \textbf{0.93}        & \textbf{70}          & \textbf{0.89}          \\ \bottomrule
\end{tabular}%

}

\vspace{-10pt}
\label{tab:small_space}

\end{table}

%% file: tables/tab_prune.tex


\begin{table}[t]
\centering
\renewcommand*{\arraystretch}{1}
\setlength{\tabcolsep}{3pt}
\caption{Pruning can speedup circuits in Pennylane.}
\vspace{-10pt}
\label{tab:prune}
\resizebox{\columnwidth}{!}{%
\begin{tabular}{lcccccccccc}
\toprule
Pruning Ratio & 0 & 0.1 & 0.2 & 0.3 & 0.4 & 0.5 & 0.6 & 0.7 & 0.8 & 0.9  \\
\midrule
Time (s) & 3.46 & 3.19 & 2.93 & 2.63 & 2.32 & 2.11 & 1.80 & 1.52 & 1.28 & 0.95 \\
Speedup & 0 & 7.9\% & 15.4\% & 24.0\% & 33.0\% & 39.1\% & 48.0\% & 55.9\% & 63.0\% & 72.4\%  \\
\bottomrule
\end{tabular}%
}
\vspace{-15pt}
\end{table}

%% file: texts/3_related.tex
\section{Related Work}
\label{sec:related}
\noindent\textbf{Quantum Machine Learning.}
Quantum machine learning (QML)~\cite{biamonte2017quantum, lloyd2020quantum, lloyd2016quantum, lloyd2014quantum, havlivcek2019supervised, liang2021can, wang2021exploration, wang2021roqnn}
explores the training and evaluation of ML models on quantum devices.
They have been shown to have potential speed-up over their classical counterparts in various tasks, including metric learning \cite{lloyd2020quantum}, data analysis\cite{lloyd2016quantum}, and principal component analysis\cite{lloyd2014quantum}. 
In modern designs, QML models use variational quantum circuits with trainable parameters -- quantum neural networks (QNNs). 
Various theoretical formulations for QNN have been proposed, e.g., quantum classifier\cite{farhi2018classification}, quantum convolution \cite{henderson2020quanvolutional}, and quantum Boltzmann machine\cite{amin2018quantum}, etc. Most prior works are exploratory and rely on classical simulation of small quantum systems\cite{farhi2018classification}. Several works also propose to search circuits~\cite{zhang2020differentiable, zhang2021neural, du2020quantum, lu2020markovian} but they neither perform noise-adaptive co-search of the circuit and qubit mapping nor have extensive evaluations on real QC devices as in \name.

\noindent\textbf{Noise-Adaptive Quantum Compiling.}
A quantum compiler translates a quantum program written in high-level programming languages to hardware instructions, which is an analogy to compiler and EDA tools in classical computations~\cite{10.5555/1177220, 9218757, wang2018learning, mao2019park, 8457638}. For NISQ systems \cite{preskill2018quantum}, such translation needs to be noise-adaptive. As such, 
Many noise-adaptive quantum compilers have been proposed. 
For example, various gate errors can be suppressed by dynamical decoupling \cite{ hahn1950spin, viola1999dynamical, biercuk2009optimized, lidar2014review}, composite pulses\cite{merrill2014progress, low2014optimal, brown2004arbitrarily}, randomized compiling\cite{wallman2016noise}, hidden inverses\cite{zhang2021hidden}, qubit mapping\cite{murali2019noise, tannu2019not, li2019tackling}, instruction scheduling\cite{murali2020software, wu2020tilt}, and frequency tuning\cite{versluis2017scalable, helmer2009cavity, ding2020systematic}. 
Typically, the key to these techniques is to find opportunities for local error cancellation within a quantum circuit. 
Instead, we propose to search for a quantum circuit and its qubit mapping pair that is the most resilient to noise. 
The flexibility in changing the quantum circuit itself gives us more freedom to build robustness into the quantum algorithms.

\noindent\textbf{Quantum Simulation.}
Beyond ML, variational circuits can also be used to explore challenging quantum many-body physics problems. 
The first implementation of variational circuits was the Variational Quantum Eigensolver (VQE)\cite{peruzzo2014variational,mcclean2016theory, kandala2017hardware} for quantum simulation of physical systems.

Prior work showed that finding such an ansatz and learning their parameters is challenging. 
Different classes of ansatz designs have been proposed: (1) \emph{Problem ansatz}\cite{peruzzo2014variational,o2016scalable} is adapted to a target problem. 
E.g., UCCSD ansatz \cite{bartlett2007coupled} is a design based on the structures in a quantum system using computational chemistry models. 
(2) \emph{Hardware ansatz}\cite{kandala2017hardware} is adapted to the properties of the computing hardware. 
Problem ansatz is shown to be typically more resilient to barren plateau than hardware ansatz\cite{mcclean2018barren}. 
In this work, our \name aims to find a balanced and robust ansatz design via SuperCircuit-based search.

 \noindent\textbf{Quantum Error Mitigation.}
As the error forms the bottleneck of the quantum area~\cite{10.5555/3433701.3433762}. Various error mitigation methods are proposed. Extrapolation methods~\cite{temme2017error, li2017efficient} perform multiple measurements under different error rates and extrapolate ideal measurement outcomes when no noise. Quasi-probability~\cite{temme2017error, huo2021self} probabilistically inserts X/Y/Z gates then sum results together to cancel out noise. Ensemble of diverse mapping~\cite{10.5555/3433701.3433720} runs different mappings of the same circuit on different machines and combines the results. Other methods such as quantum subspace expansion~\cite{mcclean2017hybrid} and learning-based mitigation~\cite{strikis2020learning, czarnik2020error} are also proposed. \name is fundamentally different. Prior work focuses on the low-level numerical correction of trained circuits; \name embraces much larger optimization freedom by searching ansatz with intrinsic robustness and performs pruning during training. The existing noise mitigation can be combined with \name as they are orthogonal.

%% file: texts/7_conclusion.tex
\section{Conclusion}

\label{sec:conclusion}
We propose \name, a noise-adaptive co-search framework for robust variational circuit and qubit mapping. We leverage the \supercircuit-based search to explore an ample design space efficiently. Iterative pruning is further leveraged to remove redundant gates in the searched circuits. Extensive experiments on QML and VQE tasks demonstrate the higher robustness and performance of \name searched circuits over baseline designs. 
We also open-source our circuit training library \quantumengine, serving as a convenient infrastructure for future research of variational quantum algorithms.

\textbf{Outlook.} Our results suggest a variety of new avenues for further theoretical and experimental explorations in the domain of variational quantum algorithms. For example: (1) \emph{Machine Learning:} How to deploy a noise-adaptive search strategy to automate the design of a quantum feature map (i.e., embedding data in high-dimensional Hilbert space)? (2) \emph{Optimization:} Can a searched variational ansatz be designed to alleviate the barren plateau issue as seen in many existing shallow ansatz? (3) \emph{Chemistry:} What are the applications amenable to the ansatz search strategy and how to use the searched ansatz to efficiently prepare low-energy eigenstates of a many-body quantum system?

\textbf{Data Availability.} Our open-source software can be found in this link: \href{https://github.com/mit-han-lab/torchquantum}{\quantumengine}. Source data will also be made available to facilitate further research.

%% file: texts/acknowledgment.tex
\section*{Acknowledgment}

We thank National Science Foundation, MIT-IBM Watson AI Lab, and Qualcomm Innovation Fellowship for supporting this research. This work is funded in part by EPiQC, an NSF Expedition in Computing, under grants CCF-1730082/1730449; in part by STAQ under grant NSF Phy-1818914; in part by DOE grants DE-SC0020289 and DE-SC0020331; and in part by NSF OMA-2016136 and the Q-NEXT DOE NQI Center. We acknowledge the use of IBM Quantum services for this work. 

%% file: texts/artifact.tex
\appendix
\section{Artifact Appendix}

\subsection{Abstract}

Our paper introduces a pipeline to search for the most robust architecture (ansatz) and qubit mapping of parameterized quantum circuits with noise information in the loop. 

The artifact contains two parts. The first is a general parameterized quantum circuit training framework called \quantumengine. It supports the contribution of our paper on a convenient framework for parameterized quantum circuit research. It can be validated by running the training of an example circuit such as a quantum neural network (QNN) to perform image classification. The second part is the \name pipeline. It supports our contribution on a method to find the most robust circuit ansatz and corresponding qubit mapping on the target quantum device. We provide scripts and Google Colab notebooks to reproduce each step of the pipeline, from \supercircuit training, noise-aware search, \subcircuit training, to pruning, and finally evaluation on real quantum hardware (IBMQ quantum machines). The results can be validated by the better performance of the searched circuits, such as the higher classification accuracy of QNN and the lower expectation value of VQE. The minimal hardware requirements will be Intel CPU, one Nvidia GPU, and remote access to IBMQ quantum machines (using our provided token). The minimal software requirements will be python libraries such as PyTorch, Qiskit, etc.




\subsection{Artifact check-list (meta-information)}


{\small
\begin{itemize}
  \item {\bf Algorithm: } SuperCircuit-based quantum circuit search. It contains:
  \begin{itemize}
      \item \supercircuit training.
      \item \subcircuit and qubit mapping co-search.
      \item \subcircuit training.
      \item pruning.
  \end{itemize}
  \item {\bf Program: } Benchmarks: 
  \begin{itemize}
      \item Image classification with quantum neural networks.
      \item Vowel recognition with quantum neural networks.
      \item Variational quantum eigensolver (VQE) task with parameterized quantum circuits.
  \end{itemize}
  \item {\bf Compilation: } Qiskit Compiler. Public available. Version 0.31.0.
  \item {\bf Binary: } PyTorch binary checkpoint files are included.
  \item {\bf Model: } Automatically searched and human designed quantum neural networks, variational quantum circuits.
  \item {\bf Data set: } 
  \begin{itemize}
      \item Vowel recognition. Public available: \url{https://www.openml.org/d/58}. Approximate size: 100KB.
      \item MNIST image classification. Public available: \url{http://yann.lecun.com/exdb/mnist/}. Approximate size: 15MB.
      \item Fashion MNIST image classification. Public available: \url{https://github.com/zalandoresearch/fashion-mnist}. Approximate size: 15MB.
  \end{itemize}
  The dataset can also be automatically accessed through our \quantumengine library with no need for explicit downloading and installing.
  \item {\bf Run-time environment: } 
  \begin{itemize}
      \item Not OS-specific.
      \item Main software dependencies: Python, PyTorch, Qiskit
      \item No need for root access.
  \end{itemize}
  \item {\bf Hardware: }
  
  \begin{itemize}
      \item Intel CPUs.
      \item Remote access to IBMQ quantum machines. Public available: https://quantum-computing.ibm.com/
      \item Nvidia GPUs for faster training.
  \end{itemize}
  \item {\bf Run-time state: } Sensitive to the noise characteristic of quantum machines. The noise on real quantum machines changes constantly, so the searched noise-aware circuit architecture and the measured performance may change.
  \item {\bf Execution: } Specific conditions during experiments: the searched circuits need to be evaluated on the real machine right after searching, otherwise the noise characteristics will drift, and performance will be degraded. The runtime depends on the available CPU/GPU machines and the size of the quantum circuits. The approximate runtime on CPU machines for searching and testing one 4-qubit quantum circuit will be around 5 to 10 hours. The approximate runtime on one Nvidia GPU will be about 5 to 10 times faster.
  \item {\bf Metrics: } 
    \begin{itemize}
        \item Classification task: top1 accuracy, the higher the better.
        \item VQE task: expectation value of molecule ground state energy, the lower the better.
    \end{itemize}
  \item {\bf Output: }
  \begin{itemize}
    \item Classification task: classification labels.
    \item VQE task: expectation value of molecule ground state energy.
\end{itemize}
  \item {\bf Experiments: } We provide shell scripts for each step of the \name pipeline. You may find the detailed instructions in \url{https://github.com/mit-han-lab/torchquantum/artifact/README.md}. The maximum allowable variation of classification accuracy should be smaller than 10\%.
  
  \item {\bf How much disk space required (approximately)?: } $<$ 10 GB
  \item {\bf How much time is needed to prepare workflow (approximately)?: } 1 hour.
  \item {\bf How much time is needed to complete experiments (approximately)?: } 24 hours.
  \item {\bf Publicly available?: } Yes. \url{https://github.com/mit-han-lab/torchquantum} and \url{https://zenodo.org/record/5787244#.YbunmBPMJhE}.
  \item {\bf Code licenses (if publicly available)?: } MIT License.
  \item {\bf Data licenses (if publicly available)?: } 
  \begin{itemize}
      \item Vowel: Creative Commons Public available.
      \item MNIST:  Creative Commons Attribution-Share Alike 3.0.
      \item Fashion MNIST: MIT License.
  \end{itemize}
  
  \item {\bf Workflow framework used?: } PyTorch, Qiskit.
  \item {\bf Archived (provide DOI)?: } \url{https://github.com/mit-han-lab/torchquantum}. We will also provide a Zenodo link and DOI at the end of the evaluation.
\end{itemize}
}

\subsection{Description}

\subsubsection{How to access}

The artifact is available at the following link:
\begin{itemize}
    \item \url{https://github.com/mit-han-lab/torchquantum}
\end{itemize}

\subsubsection{Hardware dependencies}
In order to complete the \name pipeline experiments in a reasonable amount of time, a CPU with at least 16 GB memory is necessary. Remote access to IBMQ quantum machines is necessary. We provide tokens in the artifact to access the IBMQ machine so no additional effort is required to get access. One GPU machine is highly recommended as it will significantly accelerate the training process.

\subsubsection{Software dependencies}
The artifact is implemented in Python and requires several packages such as PyTorch and Qiskit. The detailed full list of the required packages is in \texttt{requirements.txt} file and can be installed automatically when installing the \quantumengine library.

\subsubsection{Data sets}

Machine learning datasets include Vowel recognition, MNIST, and FashionMNIST.
VQE benchmarks include various molecules.
Details can be found in Table~\ref{tab:benchmarks} of the paper.

\subsubsection{Models}
The QNN models are searched \subcircuits with our noise-aware search pipeline. We also have human design, random search, no-unaware searched models as baselines.

\subsection{Installation}

\noindent You can first download the repo to local machine. Then enter the folder and run:


\noindent \texttt{pip install --editable .}

See our README.md for detailed installation instructions.

\subsection{Experiment workflow}
We provide multiple examples to run our artifact.

\subsubsection{Running the \quantumengine library}
We provide the script to construct, train and deploy a simple QNN for the MNIST task in the \texttt{artifact/example1} folder. The command is:
\texttt{./example1/1\_train\_qnn.sh}

\subsubsection{Running the \name pipeline}

(i) Run the \supercircuit training step:
\texttt{./example2/quantumnas/1\_train\_supercircuit.sh}

(ii) Run the \subcircuit and qubit mapping co-search step:
\texttt{./example2/quantumnas/2\_search.sh}

(iii) Run the \subcircuit training step:
\texttt{./example2/quantumnas/3\_train\_subcircuit.sh}

(iv) Run the pruning of searched \subcircuit:
\texttt{./example2/quantumnas/4\_prune.sh}

(v) Run the \subcircuit evaluation step:
\texttt{./example2/quantumnas/5\_eval.sh}

Furthermore, we provide scripts to train and evaluate the human baseline designs:

(i) Run the human baseline training step:
\texttt{./example2/human/1\_train.sh}

(ii) Run the human baseline evaluation step:
\texttt{./example2/human/2\_eval.sh}

We also provide other scripts for training \name in different datasets, design spaces, and quantum machines and other baselines. Please check the README.md file for the details.

\subsection{Evaluation and expected results}

For the evaluation of the functionality of \name, the quantum circuit should be successfully constructed and trained. The expected results will be decreasing training loss and increasing training accuracy on the classification task.

For the evaluation of \name, the evaluated accuracy on real quantum machines will be obtained after running the scripts. The expected results will be (1) \name searched model has better accuracy than baseline models for classification tasks. (2) \name searched model achieves lower expectation value of molecule ground state energy for the VQE tasks. This can verify the critical results of the paper in Figure~\ref{fig:devices}, as the main contribution is finding a robust quantum circuit and its qubit mapping for the real quantum devices.


\subsection{Experiment customization}
The general \quantumengine can be used to construct, train and deploy different architectures of the quantum circuits. The users can build different customized circuits from scratch.

\subsection{Notes}
Since most of the results in our paper are evaluated on real quantum machines provided by IBMQ, some of the machines such as \texttt{IBMQ\_Yorktown}, \texttt{IBMQ\_Athens}, \texttt{IBMQ\_Melbourne} have already retired, so we are only able to reproduce results on the remaining accessible quantum machines.

\subsection{Methodology}

Submission, reviewing and badging methodology:

\begin{itemize}
  \item \url{https://www.acm.org/publications/policies/artifact-review-badging}
  \item \url{http://cTuning.org/ae/submission-20201122.html}
  \item \url{http://cTuning.org/ae/reviewing-20201122.html}
\end{itemize}
